# Laser-Induced Fluorescence for measuring salt mass fraction in evaporating saline solutions

Antoine Dumain[1,2], Florine Giraud [1], Ophélie Caballina [2], Benoit Stutz [1], Guillaume Castanet [2*]

1: University of Savoie Mont Blanc, CNRS, LOCIE, 73000 Chambéry, France

2: University of Lorraine, CNRS, LEMTA, F-54000 Nancy, France

* Correspondent author: Guillaume.castanet@univ-lorraine.fr

ABSTRACT

A non-intrusive optical diagnostic based on Laser-Induced Fluorescence (LIF) is developed for measuring salt mass fraction in highly concentrated aqueous solutions relevant to absorption heat pump applications. The method relies on a systematic redshift of the fluorescence emission spectra of several dyes, likely induced by an increase of ionic strength with salt mass fraction. A two-color ratiometric approach, based on the ratio of fluorescence intensities in two spectral bands, enables robust mass fraction measurements. This study demonstrates that the techniques is not limited to LiBr but also NaCl, relevant for desalination applications. In parallel, fluorescence lifetime measurements using Time-Correlated Single Photon Counting (TCSPC) reveals that fluorescence lifetime is also affected by the LiBr mass fraction, allowing an alternative measurement technique. For both techniques, many dyes are tested and are compared in terms of salt mass fraction sensitivity. Both techniques are validated independently through comparison with shadowgraph-based measurements during evaporation of droplets. Good agreement is obtained in liquid-phase conditions over a wide range of salt mass fractions, confirming the reliability of the proposed diagnostics. Key sources of uncertainty including temperature, reabsorption effects and phase changes has been identified and analyzed. Finally, preliminary results demonstrate the potential extension of the approach toward planar imaging of evaporating droplets and more complex multiphase flows. The methodology provides a basis for future quantitative diagnostics of heat and mass transfer in desorption systems such as falling-films.

## 1. Introduction

Saline aqueous solutions are involved in many industrial processes, including absorption heat pumps, seawater desalination, drying technologies, and chemical processing. Among these applications, absorption heat pumps using lithium bromide (LiBr)/water mixtures have received growing attention because they enable the recovery and valorisation of low-grade waste heat from industrial processes. In such systems, the desorber plays a central role by evaporating water from the solution in order to increase the salt mass fraction of the working fluid. Since heat and mass transfer mainly occur at liquid–gas interfaces, the development of efficient contacting configurations is of primary importance. While falling liquid films over heated surfaces currently represent the most common technological solution, droplets and sprays are increasingly considered because of their potentially enhanced transfer performance (Altamirano et al., 2020).

Saline evaporation processes are also encountered in seawater desalination, where salts such as sodium chloride (NaCl) must be separated from water through membrane or thermal processes (Al-Obaidi et al., 2025), as well as in drying technologies used for the production of powders, pharmaceuticals, and chemicals. Many of these processes involve rapidly evolving liquid–gas interfaces, including evaporating droplets, sprays or falling films, for which local concentration measurements remain particularly challenging.

In many of these situations, local measurements of salt concentration are highly attractive because they preserve the hydrodynamics of the flow, but they remain particularly challenging because of the transient nature of the flows and the presence of rapidly evolving liquid–gas interfaces. Raman spectroscopy is a well-established technique for composition measurements. For instance, quantitative analysis of Raman OH stretching bands has been used to determine NaCl concentration in fluid inclusions based on salinity-dependent variations of spectral sub-bands (He et al., 2024; Mernagh & Wilde, 1989). However, extending these approaches toward fast time-resolved measurements remains difficult because of the weak intensity of the Raman signal and the need for detailed spectral analysis. Rainbow refractometry has also proven to be an elegant and efficient technique for measuring droplet temperature and composition when the refractive index depends on temperature or concentration. In evaporating sprays, global rainbow refractometry has been adapted to monitor refractive index variations in real time (Li et al., 2020). Nevertheless, because the technique relies on light scattering by nearly spherical droplets, its applicability remains restricted to specific flow configurations and cannot easily be extended to geometries such as thin liquid films. In this context, Laser-Induced Fluorescence (LIF) appears as a promising alternative for concentration measurements in evaporating saline solutions. LIF diagnostics have already been extensively applied to liquid films (Collignon et al., 2022) and in droplet flows (Lemoine & Castanet, 2013), especially for temperature and mixing measurements. However, to the best of our knowledge, the use of LIF for salinity measurements in highly concentrated saline solutions, particularly under conditions approaching crystallization, has not yet been reported. Although fluorescent dyes such as Rhodamine B have been widely employed as temperature tracers in heat transfer and evaporation studies, fluorescence is also known to be sensitive to the chemical environment of dye molecules (Lakowicz & Masters, 2008). In particular, fluorescence techniques have long been exploited for pH measurements (Coppeta & Rogers, 1998), including applications involving $CO_2$ bubble wakes and reactive flows (Kong et al., 2018; Lacassagne et al., 2018). The sensitivity of fluorescence to liquid composition was also investigated for evaporating binary droplets composed of water mixed with alcohols or ketones (Stiti et al., 2024). In this study, fluorescence lifetime measurements were generally preferred because of their robustness with respect to excitation and collection conditions. Similar approaches were also adopted for chloride-sensitive fluorescent probes developed for seawater salinity measurements (Huber et al., 2001) and, more recently, for intracellular $Na^+$ concentration measurements (Meyer et al., 2019). By contrast, intensity-based fluorescence measurements, and especially ratiometric approaches relying on intensity ratios measured in two spectral bands, appear to have received comparatively little attention for salinity measurements. Yet, such approaches could provide a relatively simple and fast experimental methodology compatible with transient evaporation phenomena and future developments toward spatially resolved diagnostics. It should also be emphasized that fluorescence lifetime and ratiometric intensity measurements do not rely on identical physical observables. While fluorescence lifetime measurements are mainly sensitive to variations in fluorescence decay dynamics, ratiometric approaches require measurable salt-induced distortions of the fluorescence emission spectrum. Consequently, the fluorophores providing the highest sensitivity may differ depending on the selected diagnostic approach.

The present work is part of a broader project dedicated to LiBr absorption heat pumps and focuses on the development of fluorescence-based optical diagnostics for saline evaporation processes. More specifically, the objective of this first study is to establish the spectroscopic basis and evaluate the feasibility of Laser-Induced Fluorescence (LIF) measurements for determining salt concentration in highly concentrated aqueous solutions. Although the main focus is placed on LiBr solutions, additional experiments performed with NaCl suggest that similar methodologies could potentially be extended to other saline mixtures.

First, the influence of salt concentration, considering both LiBr and NaCl aqueous solutions, on the fluorescence emission spectra and fluorescence lifetime of several dyes is investigated experimentally. Based on these measurements, both lifetime-based and intensity-based approaches are evaluated for salt concentration measurements. Particular attention is devoted to a one-dye/two-color ratiometric LIF technique because of its relative simplicity and its potential compatibility with fast measurements and future spatially resolved diagnostics. The proposed diagnostic relies on a photomultiplier-based detection system providing high sensitivity and fast temporal response, and is validated on evaporating droplets. Finally, preliminary results illustrating the possible extension of the approach toward planar laser-induced fluorescence imaging are briefly discussed.

## 2. Spectral sensitivity of fluorescent dyes to salt mass fraction

### 2.1 Experimental setup and measurement procedure

The experimental setup used to characterize the salt mass fraction sensitivity of fluorescent dyes is shown in Figure 1. Fluorescence measurements were performed in a cubic quartz cuvette with a side length of 5 cm. Initial saline solutions were prepared close to saturation at 20 °C, corresponding to approximately 60 wt.% for LiBr and 25 wt.% for NaCl. Fluorescent dyes were then added to the solutions at low concentration, typically of the order of $10^{-7}$ M. Such low dye concentrations ensure negligible attenuation of both laser excitation and fluorescence emission within the optical path of the cuvette for all investigated conditions. The cuvette was equipped with a heating element, a thermocouple, and a magnetic stirrer to maintain homogeneous temperature and concentration fields during the experiments. Particular attention was devoted to temperature control because some salts, especially LiBr, release a significant dilution heat when mixed with water, and some dyes are very sensitive to temperature. The solution temperature was therefore maintained within ±0.5 °C of the target value.

Fluorescence emission spectra were recorded using Ocean Insight QE Pro or Hamamatsu spectrometers, depending on the wavelength range of the spectrum (limited to [532-700] nm for the Ocean Insight QE Pro).

After dark-noise acquisition, the fluorescent solution was illuminated by a laser beam and the emitted fluorescence was collected through a lens at 90° relative to the excitation direction connected to the spectrometer. Following each spectral acquisition, distilled water was progressively added to decrease the salt mass fraction. After thermal equilibrium was re-established, a new measurement was performed. This procedure was repeated to cover a wide range of salt mass fractions, from near-saturation conditions down to highly diluted solutions. To minimize photobleaching and dye photodegradation, the laser source was switched off between successive measurements.

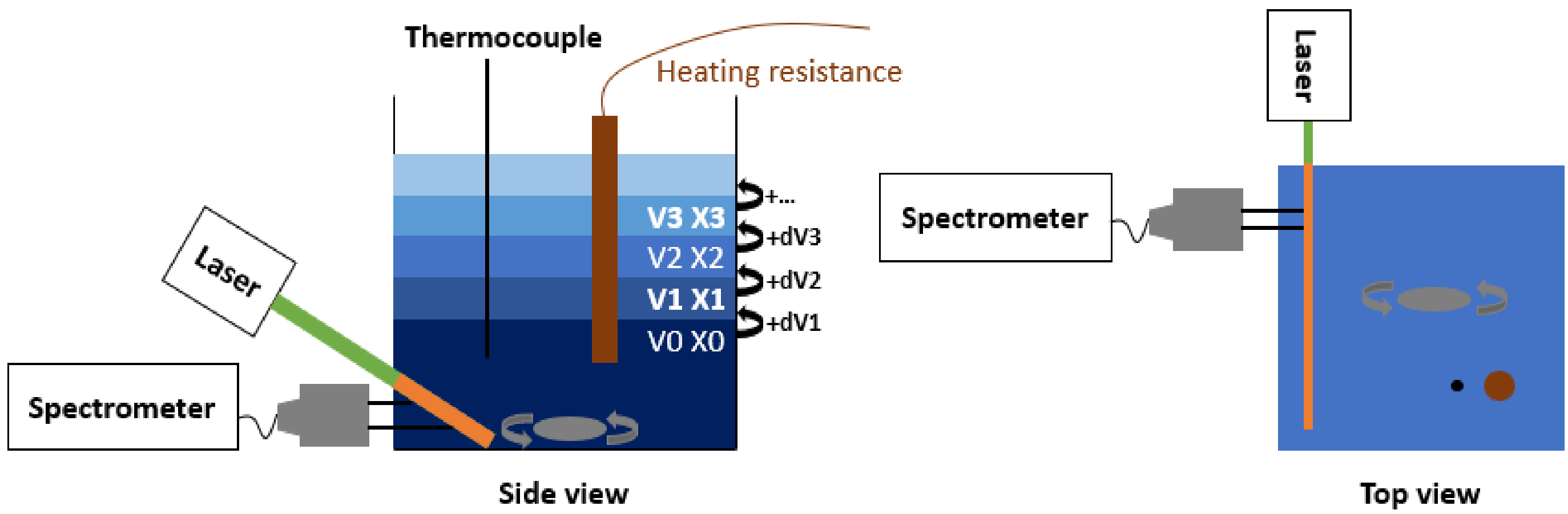

**Figure 1** : Experimental setup used for the measurement of salt mass fraction sensitivities of fluorescent dyes

In the present study, commercially available fluorescent dyes were selected because of their accessibility and compatibility with both pulsed and continuous-wave laser excitation, allowing future developments toward both pointwise and planar diagnostics. The investigated dyes include Kiton Red (KR), Rhodamine 19 (Rh19), Rhodamine 6G (Rh6G), Eosin Y (EY), Rhodamine WT (RhWT), Sulforhodamine G (SRhG), Oxazine 4 (Ox4), and Styryl 6 (St6). Most measurements were performed using a frequency-doubled Nd:YAG laser at 532 nm, except for RhWT, SRhG and EY, excited at 490 nm for limiting spectral conflicts with the laser and having sufficient amount of signal.

## 2.2 Salt-induced spectral shifts and ratiometric sensitivity

Figure 2 presents representative fluorescence emission spectra obtained for different salt mass fractions, temperatures, and fluorescent dyes. For clarity, only the spectra of KR, Rh19 and EY are shown here, while the complete dataset for all investigated dyes is provided in Appendix A. The spectra were normalized by their maximum intensity in order to emphasize spectral shifts and spectral distortions independently of absolute fluorescence intensity variations.

A remarkable and systematic result observed for all investigated dyes and experimental conditions is a salt-induced redshift of the fluorescence emission spectrum. As the salt mass fraction increases, the fluorescence peak progressively shifts toward longer wavelengths. The amplitude of this shift can become significant, reaching approximately 16 nm for KR between pure water and a 60 wt.% LiBr solution, and up to 20 nm for Ox4. This behavior likely originates from modifications of the excited-state energy levels of the dye molecules caused by increasing ionic strength, which alters solvation effects and intermolecular interactions (Stoughton & Rollefson, 1939). This shift is also observed on the absorption spectrum, available in Appendix B.

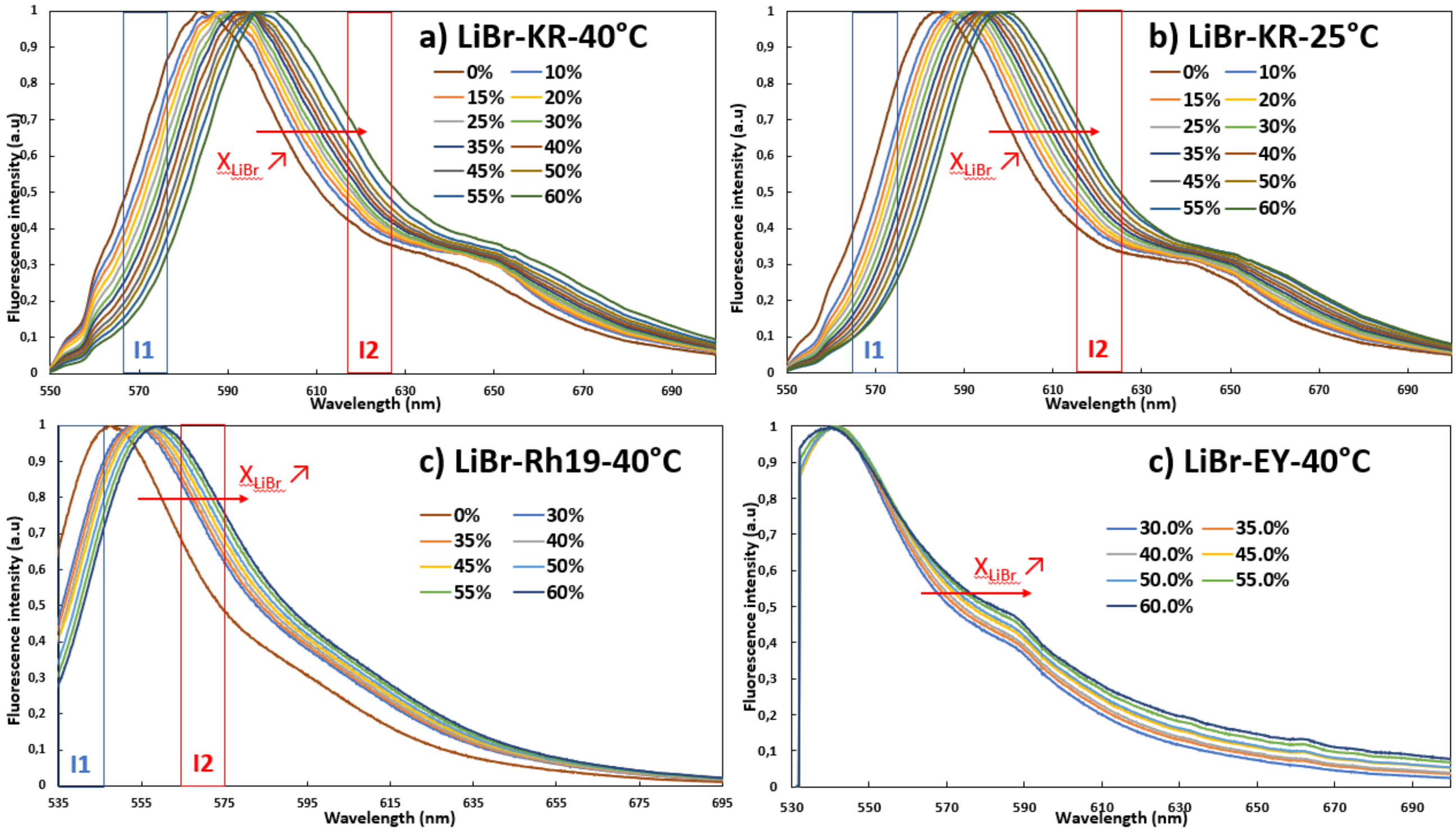

**Figure 2 :** Spectral shift with salt mass fraction for various temperature, fluorescent dye and salt species

Beyond its physical interest, this spectral redshift is particularly attractive for ratiometric fluorescence measurements. Indeed, when two spectral bands are selected on either side of the emission peak, the fluorescence intensity in the lower-wavelength band decreases with increasing salt concentration, while the intensity in the higher-wavelength band increases correspondingly. This behavior naturally enables a one-dye/two-color measurement strategy based on the ratio of fluorescence intensities measured in two spectral bands. Compared with single-band fluorescence measurements, such a ratiometric approach strongly reduces the influence of optical disturbances caused by interface deformations, variations of the measurement volume, laser intensity fluctuations, or dye concentration changes during dilution. Interestingly, unlike two-color LIF thermometry where two different dyes are often required to maximize ratio sensitivity, the salt-induced spectral deformation observed here appears sufficiently strong to allow efficient ratiometric measurements using only a single fluorescent dye.

Based on the measured emission spectra, two spectral bands denoted $I_1$ and $I_2$ were selected in order to maximize the sensitivity of the fluorescence ratio to salt concentration. Experimental observations indicate that the normalized fluorescence intensity:

$$I_n(\lambda, X) = \frac{I_{\lambda,X}}{I_{\lambda_{max},X}} \qquad \text{(Eq. 1)}$$

approximately follows an exponential evolution with salt mass fraction:

$$I_n(\lambda, X) = K \cdot \exp\left(-s_\lambda X\right) \quad \text{(Eq. 2)}$$

where X denotes the salt mass fraction and $s_\lambda$ represents the local spectral sensitivity to concentration. Following the approach proposed by Chaze et al., (2016), the sensitivity can be estimated experimentally from:

$$s_\lambda \approx \frac{\ln\left(\frac{I_n(\lambda,X_2)}{I_n(\lambda,X_1)}\right)}{X_2 - X_1} \qquad \text{(Eq. 3)}$$

Figure 3 presents the spectral evolution of $s_\lambda$ obtained for KR and Rh19 in LiBr aqueous solutions at 40 °C using X1=30 wt.% and X2=60 wt.%. Negative sensitivities are observed on the low-wavelength side of the spectrum, while positive sensitivities are obtained on the high-wavelength side because of the salt-induced redshift. The sensitivity evolves rapidly over a limited spectral range, varying for instance from approximately −2.5%/ wt. % to +1%/ wt. % within a few tens of nanometers for KR.

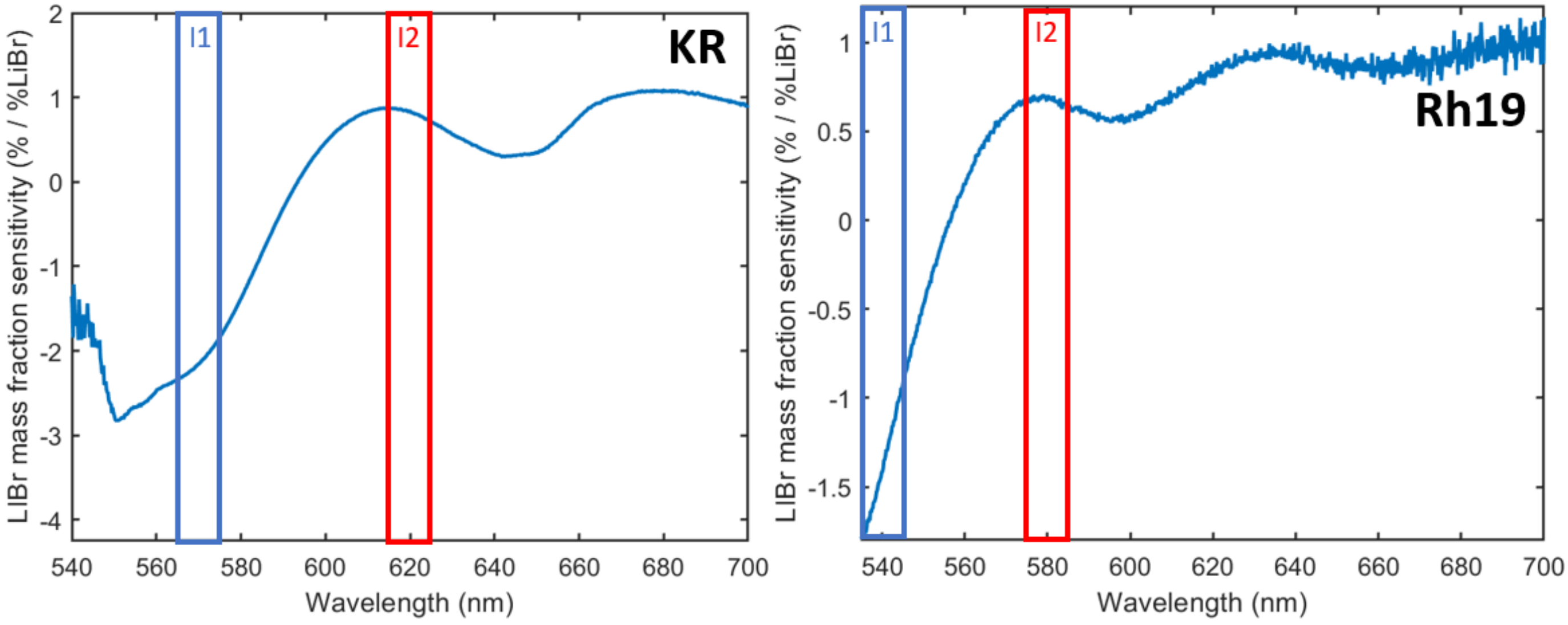

**Figure 3** : LiBr mass fraction sensitivity of Kiton Red and Rhodamine 19 at 40 °C by wavelength.

Consequently, maintaining a high sensitivity of the fluorescence ratio requires the use of relatively narrow spectral bands. A bandwidth of 10 nm was retained as a compromise between ratio sensitivity and fluorescence signal level.

Defining the fluorescence ratio R as:

$$R = I_1(X)/I_2(X) \qquad \text{(Eq.4)}$$

the previous relations lead to:

$$\mathrm{R(X)} = \mathrm{R}_0 \exp\left(-\Delta s\, X\right) \qquad \text{(Eq. 5)}$$

where $\Delta s = s_1 - s_2$corresponds to the difference between the spectral sensitivities of the two selected bands.

Interestingly, the spectral bands maximizing $\Delta s$ s generally exhibit comparable fluorescence intensities over the investigated concentration range. This feature is advantageous because it avoids the strong signal imbalance frequently encountered in two-color LIF thermometry, thereby improving practical measurement accuracy.

Figure 4 presents the logarithm of the fluorescence ratio for KR and Rh19 as a function of LiBr mass fraction. A nearly linear evolution is observed, consistent with the previous model. KR exhibits significantly larger sensitivity than Rh19, resulting in higher concentration resolution.

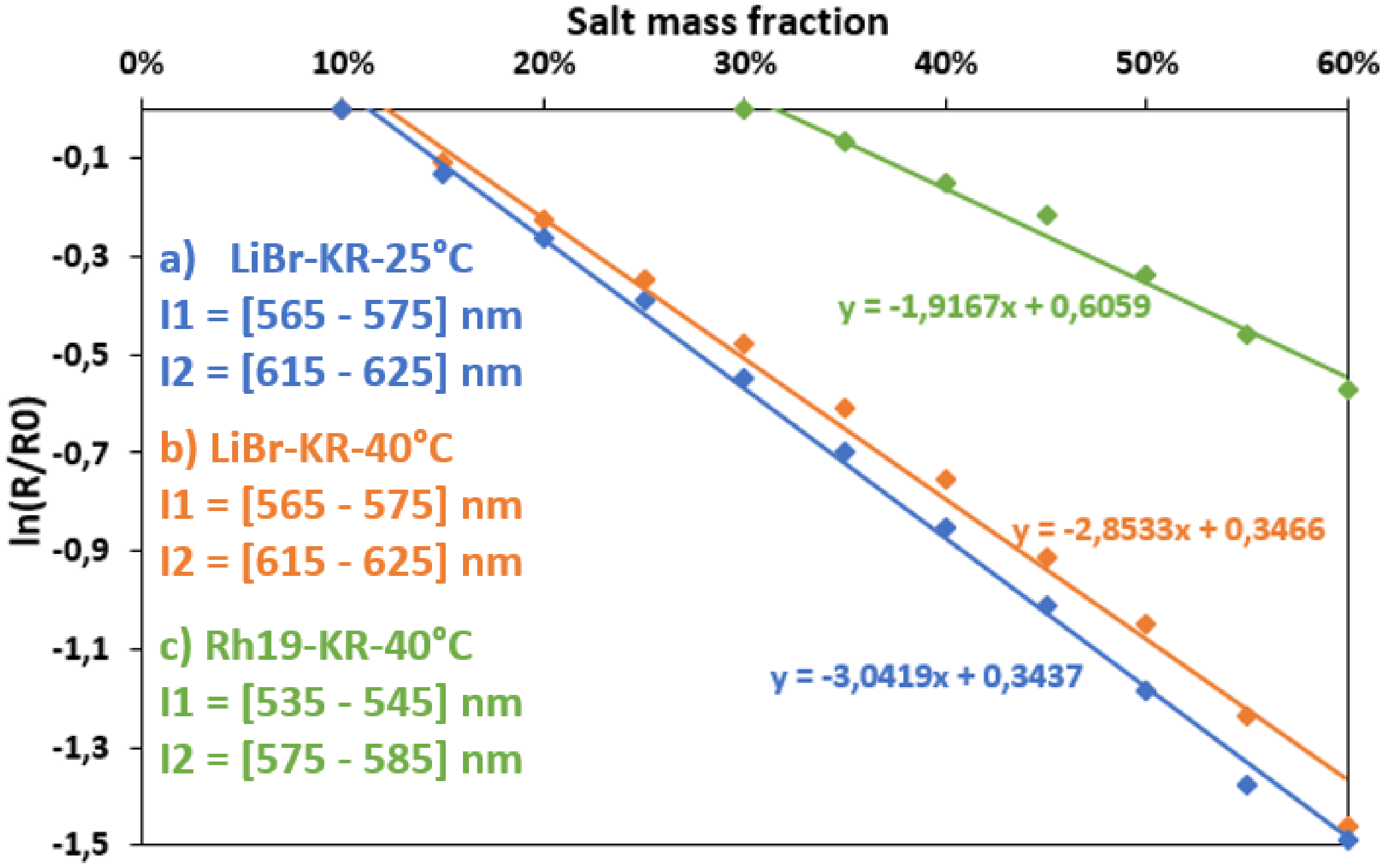

**Figure 4 :** LiBr mass fraction sensitivities. Ratio are normalized to the value R0 at lowest mass fraction X0.

Sensitivities measured for various dyes in LiBr solution at 40°C are presented in Table 1. All studied dyes have a more or less pronounced sensitivity to LiBr mass fraction. Ox 4 shows the best sensitivity with -3.477 % / % LiBr, followed by KR and RhWT. For EY, significant spectral conflict with the excitation laser was observed due to the low fluorescence emission intensity. Although the laser line was subsequently filtered using a 532 nm high-pass filter, the fluorescence emission peak remained too close to the filter cut-off wavelength. Consequently, there was insufficient spectral range available for defining the first spectral band which was therefore left empty in Table 1.

Sensitivity is not the only parameter to be taken into account for selecting an appropriate fluorescent dye. Other parameters such as the dye solubility in water, amount of emitted fluorescence light (depending on the absorption cross section at the laser wavelength and the fluorescence quantum yield of the dye), resistance to photodegradation or dependence on other external parameters such as temperature have to be carefully taken into account.

**Table 1 :** Spectral salinity sensitivity measured for various fluorescent dyes in LiBr solution at 40°C

| Dye | KR | Rh19 | Rh6G | EY | RhWt | SRhG | Ox4 | St6 |
|---|---|---|---|---|---|---|---|---|
| $\lambda_{exc}$ (nm) | 532 | 532 | 532 | 490 | 490 | 490 | 532 | 532 |
| I1 | 565-575 | 535-545 | 540-550 | - | 565-575 | 540-550 | 625-635 | 650-660 |
| I2 | 615-625 | 575-585 | 585-595 | - | 610-620 | 575-585 | 675-685 | 720-730 |
| $\Delta s$ (%/%LiBr) | -2.853 | -1.917 | -1.731 | - | -2.54 | -2.36 | -3.472 | -1.610 |

## 2.3 Influence of temperature and fluorescent dye selection

Although the proposed ratiometric approach primarily relies on salt-induced spectral distortions, the influence of temperature must also be carefully assessed because many fluorescent dyes are intrinsically temperature-sensitive. In particular, Kiton Red is well known in the LIF literature for its strong temperature dependence, which has frequently been exploited for fluorescence thermometry.

Consequently, an additional series of experiments was conducted to quantify the influence of temperature on the fluorescence ratio while maintaining the same spectral bands previously optimized for concentration measurements. The saline solution was progressively heated under a closed cover in order to limit water evaporation and therefore maintain an approximately constant salt mass fraction during the measurements.

Figure 5a presents the evolution of the fluorescence ratio measured for KR in logarithmic scale as a function of temperature. The results indicate that the ratio remains only weakly affected by temperature variations over the investigated range. This behavior can be explained by the fact that the temperature sensitivities of the two selected spectral bands remain relatively similar. Achieving a significantly smaller temperature dependence of the ratio would require selecting the lower-wavelength band further away from the emission peak, in a spectral region where fluorescence intensity becomes substantially weaker.

The corresponding error on LiBr mass fraction measurements obtained without temperature correction is shown in Figure 5b, using 20 °C as the reference calibration temperature. Although the temperature influence remains significantly smaller than the salt concentration sensitivity, it cannot be completely neglected for large temperature variations. The error remains close to 1 wt.% LiBr for temperatures below approximately 50 °C, but may reach 3–4 wt.% for highly concentrated LiBr solutions at higher temperatures. These results indicate that KR constitutes an attractive compromise for concentration measurements because of its strong salinity sensitivity, high fluorescence signal level, excellent solubility in water and good resistance to photobleaching. In addition, its emission spectrum remains sufficiently separated from the laser wavelength at 532 nm, simplifying spectral filtering in the high-sensitivity regions. Its main limitation remains its moderate temperature dependence.

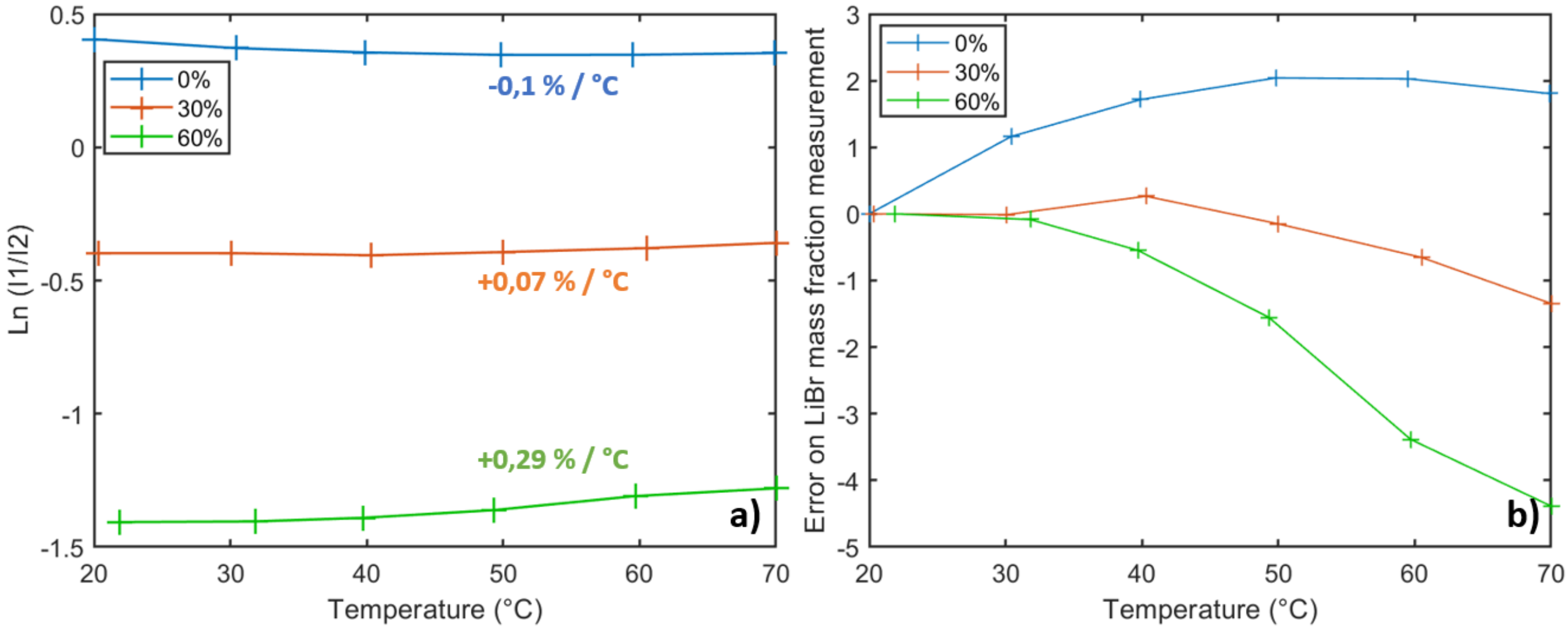

**Figure 5 :** Effect of temperature on spectral band ratio a) and on the LiBr mass fraction measurement

## 2.4 Extension to NaCl aqueous solutions

Finally, the proposed methodology was also evaluated for NaCl aqueous solutions using KR as fluorescent dye. Figure 6 presents the evolution of the fluorescence ratio in logarithm scale as a function of the NaCl mass fraction for temperatures of 25 °C and 40 °C. Spectral datasets are provided in the appendix A.

Similarly to LiBr solutions, a clear and monotonic evolution of the fluorescence ratio with salt mass fraction is observed. The measured sensitivity remains comparable to that obtained for LiBr solutions, although slightly weaker. In addition, the influence of temperature on the fluorescence ratio remains relatively moderate for low temperature variations, following trends similar to those previously observed for LiBr, as represented in Figure 7.

These results suggest that the salt-induced spectral redshift responsible for the ratiometric sensitivity is not specific to LiBr solutions and may potentially be exploited for a broader range of saline aqueous mixtures. Although the present work mainly focuses on LiBr because of its relevance for absorption heat pump applications, the methodology could therefore be extended to other saline systems encountered in desalination or drying technologies.

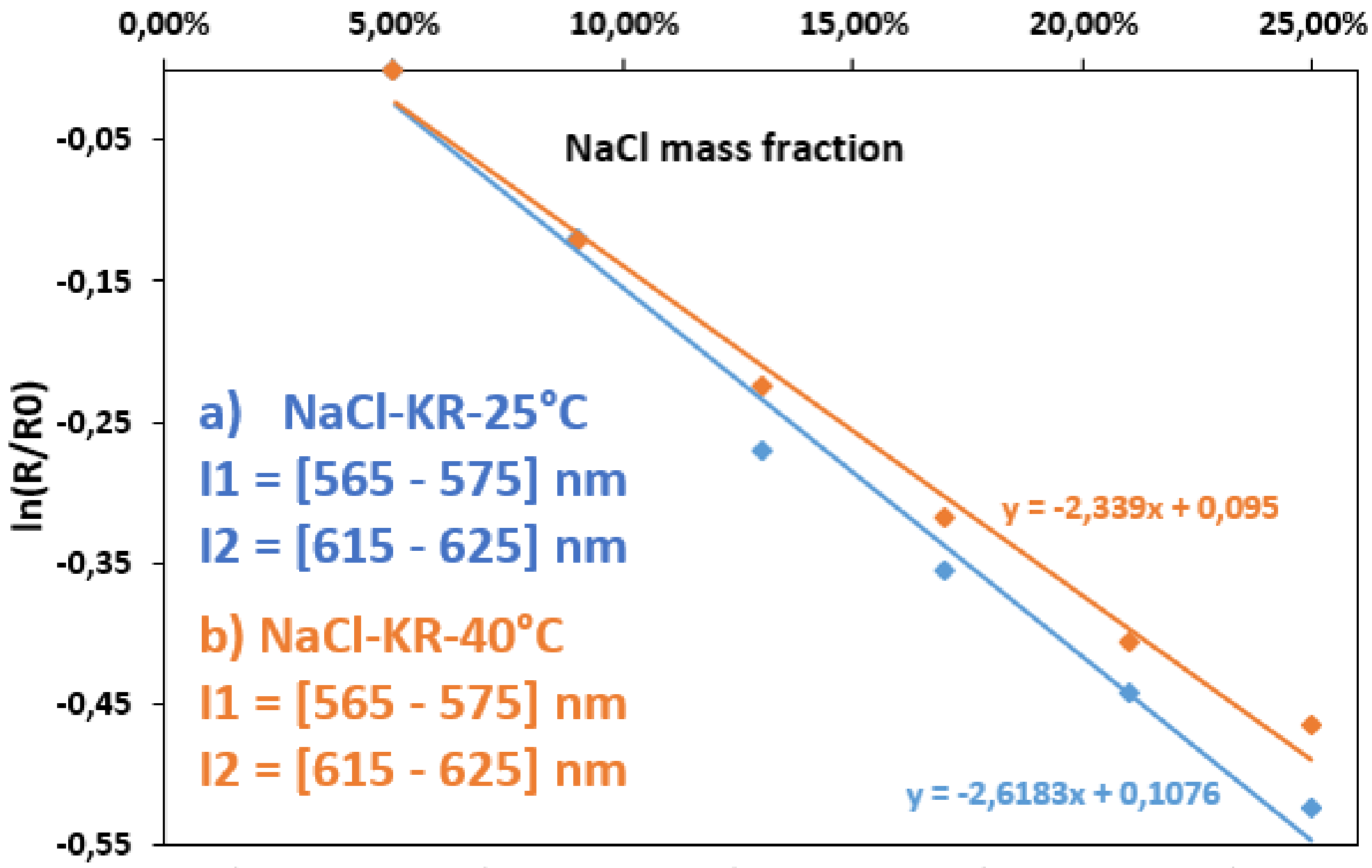

**Figure 6 :** LiBr mass fraction sensitivities. Ratio are normalized to the value R0 at lowest mass fraction X0.

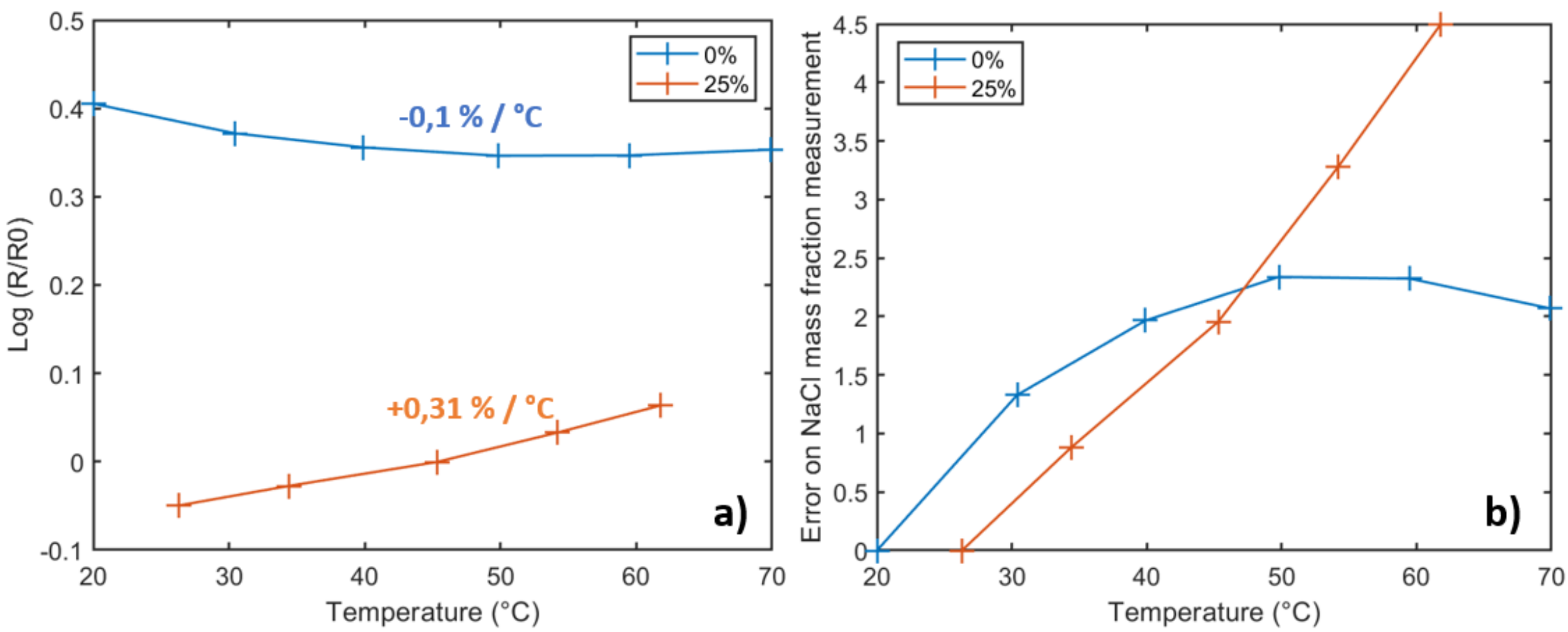

**Figure 7 :** Effect of temperature on spectral band ratio a) and on the LiBr mass fraction measurement

## 3. Study of salt mass fraction effect on fluorescence lifetime

This spectroscopic study was complemented by Time-Correlated Single-Photon Counting (TCSPC) measurements, which provide direct access to fluorescence decay dynamics. In this technique, fluorescence photons are detected individually, and their arrival times are recorded relative to the excitation pulse. Denoting $I(t)$ the fluorescence decay function, the fluorescence lifetime τ is defined as:

$$\tau = \frac{\int_0^\infty t\, I(t)\, dt}{\int_0^\infty I(t)\, dt} \quad \text{(Eq. 7)}$$

This characteristic time scale, typically on the order of a few nanoseconds, is directly related to the fluorescence quantum yield and may therefore depends on both temperature and chemical environment, including salt mass fraction.

The experimental setup is similar to that reported in previous studies elsewhere (Stiti et al., 2021) and is essentially represented in Figure 8. Excitation is provided by a femtosecond pulsed laser initially at 980 nm, and halved at 490 nm by a second harmonic generator. A fast photodiode placed after a beam splitter is used to record a fraction of the laser pulse (approximately 1%) as a timing reference. Fluorescence emission is collected using an optical lens and detected by a hybrid photomultiplier tube (PMT). The signal is then processed by a time-correlated counting module, which records the time delay between each detected photon and the excitation pulse. Further details on the setup can be found in Stiti et al., (2021). The detection wavelength range is set to 590–610 nm. Each decay curve is reconstructed from approximately $10^7$ detected photons. Same procedure as in section 2.1 is applied for solutions preparation and dilution.

Fluorescence decays obtained for Eosin Y (EY), Kiton Red (KR), Rhodamine 6G (Rh6G), and Rhodamine 19 (Rh19) in aqueous LiBr solutions at 40 °C are presented in Figure 9 for different salt mass fractions.

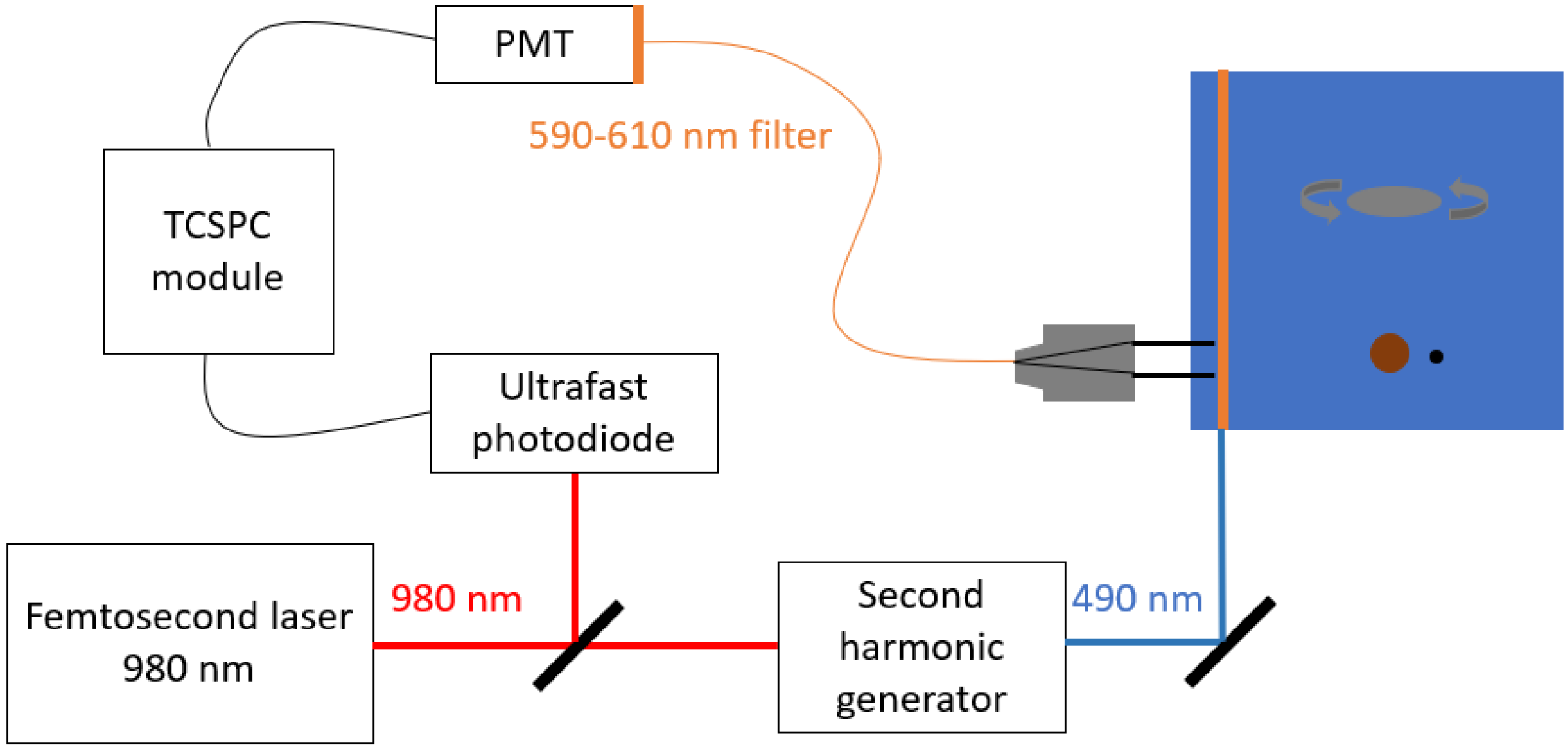

**Figure 8 :** Experimental TCSPC setup used for fluorescence lifetime calibration

For KR, the fluorescence lifetime increases as the LiBr mass fraction increases, while the opposite trend is observed for Eosin Y, Rh6G, and Rh19. Rh6g and Rh19 follow a mono exponential decay, while more complex behavior while at least biexponential decays better describe EY and KR cases.

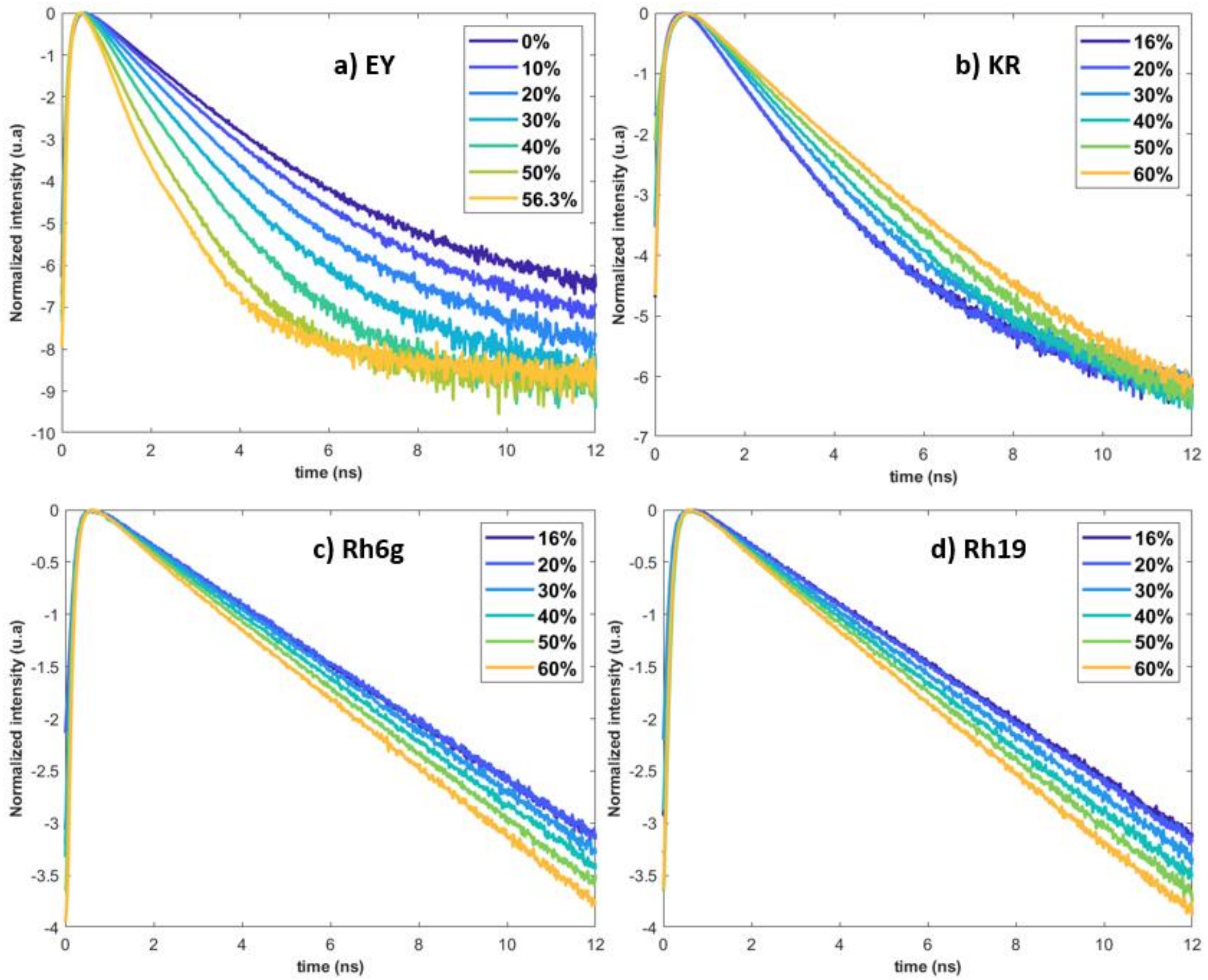

**Figure 9 :** Fluorescent decay of a) Eosyn Y, b) Kiton Red, c) Rhodamine 6g and d) Rhodamine 19 for various LiBr mass fraction at 40°C and [590-610] nm.

Figure 10 reports the normalized fluorescence lifetime as a function of LiBr mass fraction. Among the tested dyes, EY exhibits the strongest sensitivity to salt concentration, followed by Rh19 and KR. This behavior is consistent with previous studies in which EY was shown to be highly sensitive to solvent polarity variations (Stiti et al., 2024), while maintaining a relatively weak sensitivity to temperature in various solvent including water.

The temperature dependence of Eosin Y fluorescence lifetime is further investigated in Figure 11 for three salt mass fractions (0 wt.% taken from Stiti et al, 2024, 30 wt.% and 60 wt.% LiBr). No clear trend is observed for the effect of the temperature on fluorescence lifetime, that remains close to the experimental uncertainty (approximately 0.1-0.15 ns). This indicates that within the investigated range, the fluorescence lifetime of EY is significantly more sensitive to salt concentration than to temperature.

These dyes were also tested in NaCl aqueous solutions and fluorescence lifetime remained constant over the whole range of investigated salt mass fraction (0 wt.% - 25wt. %).

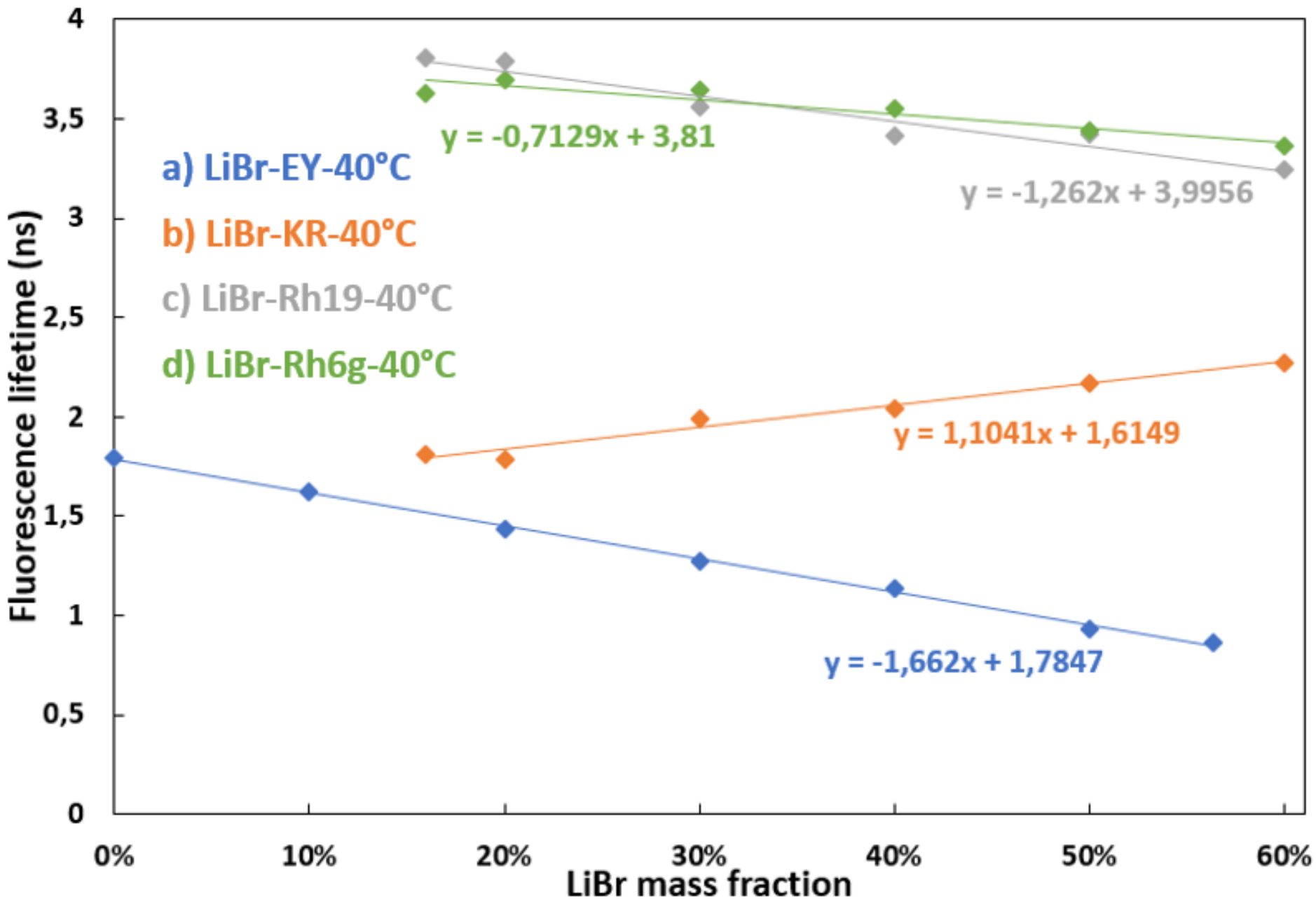

**Figure 10 :** Calibration curve for the fluorescent decay of EY, KR, Rh6g and Rh19 with LiBr mass fraction at 40 °C

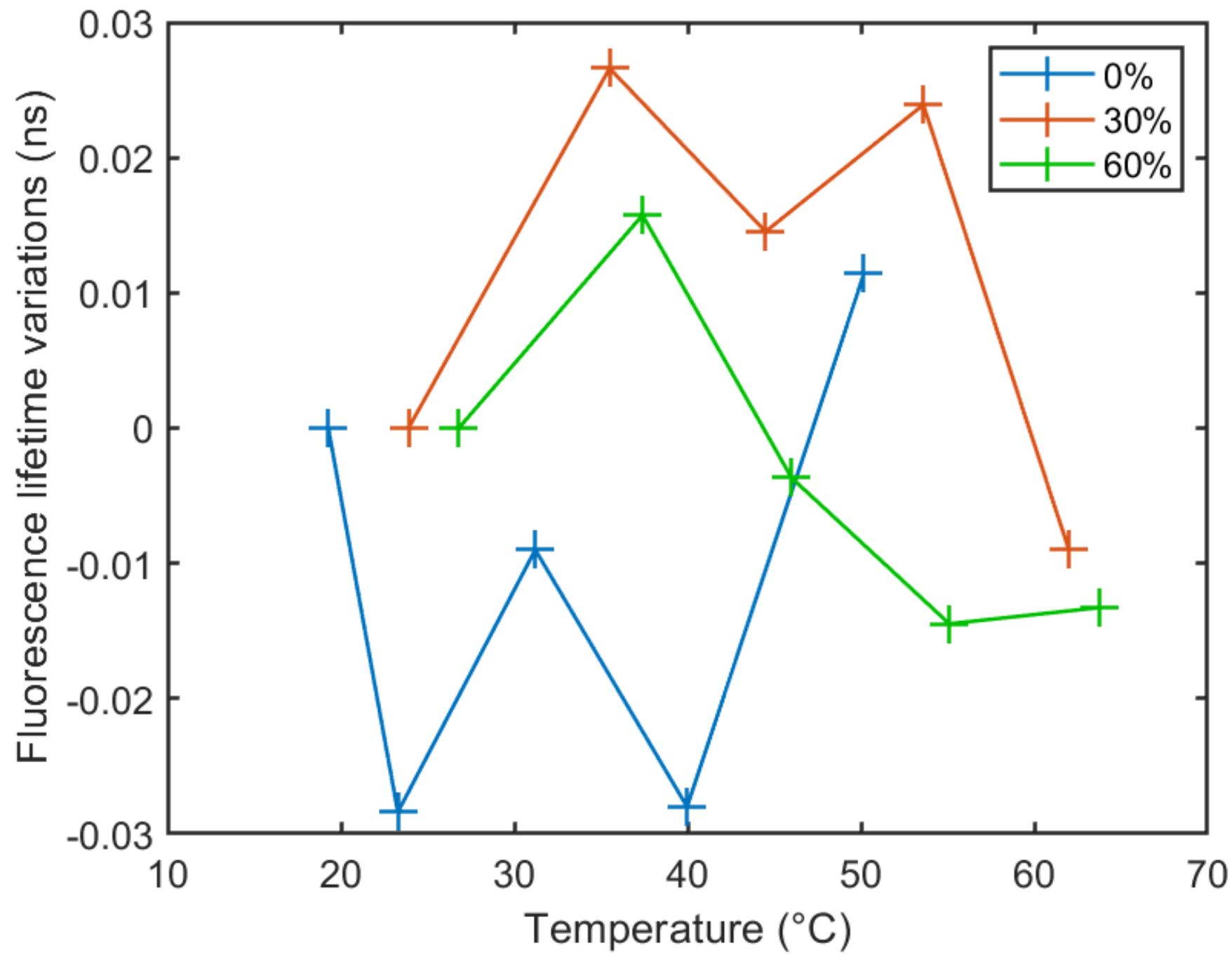

**Figure 11 :** a) Evolution of the fluorescence lifetime of EY in 0 wt.%, 30 wt.% and 60 wt.% LiBr with temperature

These results highlight that fluorescence lifetime and intensity-based ratiometric measurements probe different physical mechanisms, leading to different optimal dye selections depending on the targeted diagnostic.

## 4. Validation of LIF techniques applied on evaporating droplets

### 4.1 Evaporation of a sessile heated droplet: intensity-based LIF and shadowgraphy

To validate the proposed fluorescence-based measurement approach, the evaporation of a sessile droplet deposited on a heated sapphire substrate was investigated. The LIF technique was combined with shadowgraph imaging, providing two independent methods for estimating the evolution of the salt mass fraction during evaporation.

The experimental setup is shown in Figure 12. The sapphire substrate is mounted in a steel holder equipped with four cartridge heaters ensuring homogeneous heating. Thermocouple fixed on the upper surface allow closed-loop temperature control via adjustment of the heater power supply and temperature measurement. Once thermal equilibrium is reached, a droplet of approximately 10 µL containing an initially low LiBr mass fraction is gently deposited on the sapphire surface.

**LIF ratiometric measurement system**

Fluorescence measurements are performed using a fiber-coupled optical probe. A continuous-wave 532 nm laser (Power 2W) is delivered through an optical fiber into a collimation module, producing an expanded beam. The excitation beam is reflected by a 552 nm dichroic mirror and focused onto the droplet using an achromatic triplet lens (f = 40 mm), resulting in a measurement spot of approximately 0.23 mm. Fluorescence emission is collected through the same optical path, filtered using a 532 nm long-pass filter to reject scattered laser light, and directed toward two photomultiplier tubes equipped with appropriate spectral filters. The acquisition rate is $10^6$ Hz, with effective temporal averaging resulting in a final sampling frequency of 1 kHz. The optical configuration was verified to preserve the same spectral sensitivity to LiBr concentration as the spectrometer-based setup described in Section 2. Achromatic collimators (RC02SMA-P01) are used to minimize chromatic aberrations due to optical components. Laser light is turned off except during the measurement (few seconds) to minimize photobleaching and photodegradation.

**Shadowgraph-based mass fraction measurement**

In parallel, droplet evaporation is also monitored using shadowgraph imaging. The droplet is back-illuminated by a LED panel, and images are recorded throughout the evaporation process using a camera. Image processing consists of background subtraction, rotation to align the substrate horizontally, and binarization. The solid substrate is masked out, and the droplet contour is extracted. The endpoints $(x_1, y_1)$ and $(x_2, y_2)$ and apex position $(x_c, y_c)$ are identified. Assuming axisymmetry, the droplet volume is reconstructed by integrating stacked disks. The salt mass fraction is then obtained from mass conservation:

$$X_{LiBr} = \frac{m_{LiBr}}{V * \rho(X_{LiBr})} \quad \text{(Eq. 6)}$$

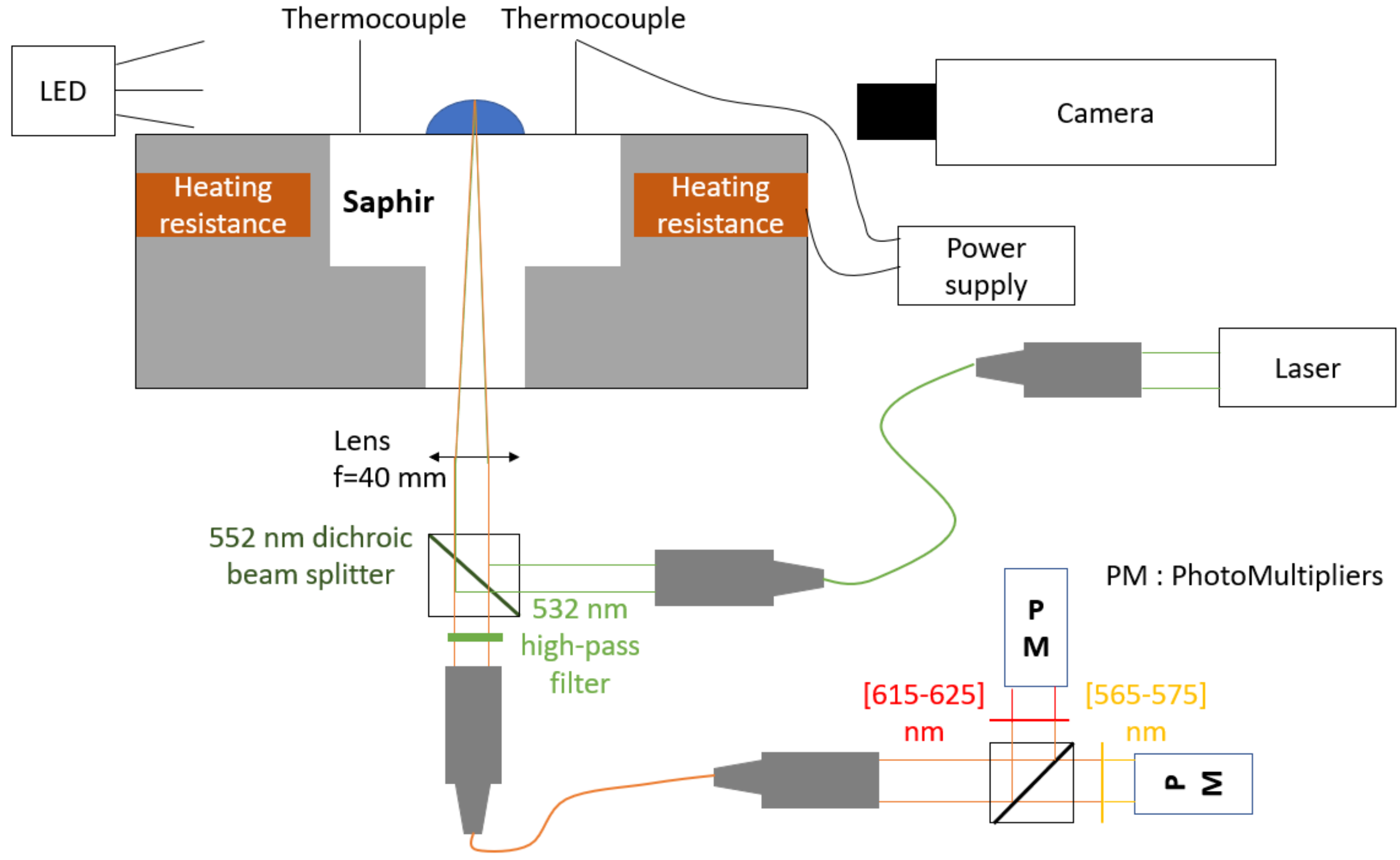

**Figure 12 :** Experimental setup for validating the intensity-based technique on a sessile droplet

where the density is approximated using an empirical correlation fitted from CoolProp 7.2.0 data (Bell et al., 2014):

$$\rho(X_{LiBr}) = 1.0111\, X_{LiBr}^2 + 0.5605\, X_{LiBr} + 1.0037 \qquad \text{(Eq. 7)}$$

In the same way, a linear density model is used for NaCl:

$$\rho(X_{NaCl}) = 0.0074\, X_{NaCl} + 0.9901 \qquad \text{(Eq. 8)}$$

A representative evaporation sequence for a LiBr droplet is shown in Figure 12, including raw images and processed binary contours used for volume reconstruction.

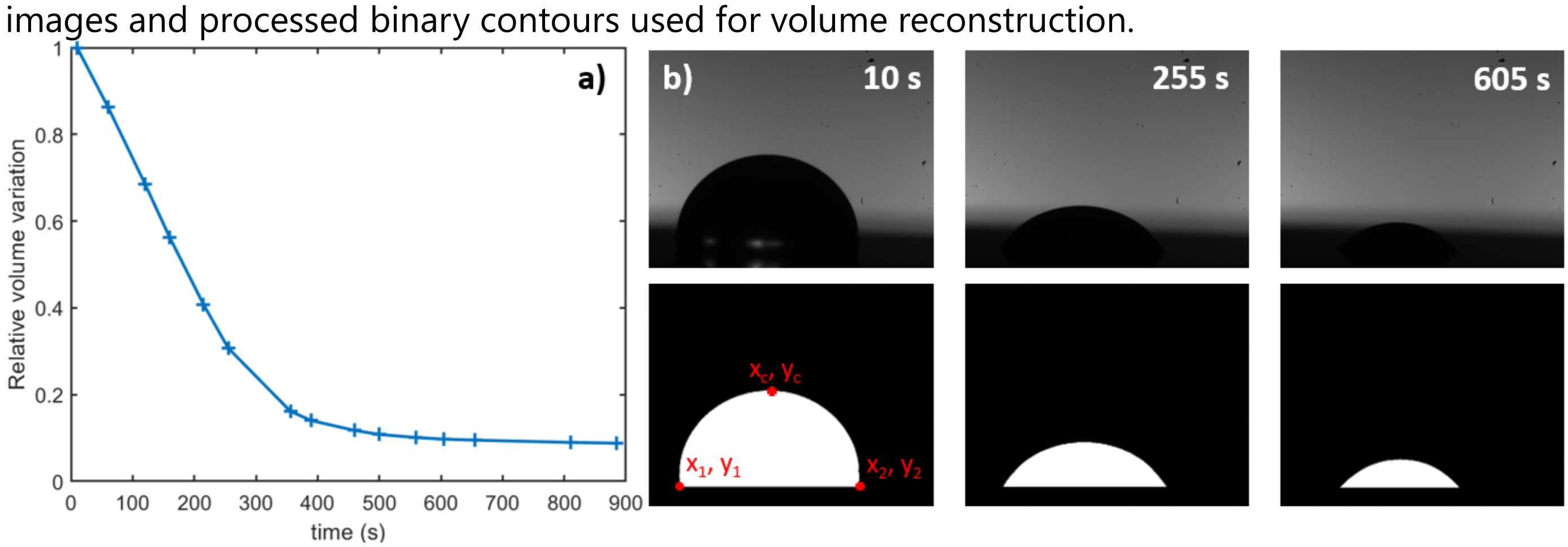

**Figure 13 :** Typical volume reduction of a sessile LiBr droplet a) and images taken (top row), and images after treatment (bottom row). X0 = 10 %, Ts = 40 °C

### Comparison between LIF and shadowgraph measurements

Figure 14 compares the evolution of LiBr mass fraction obtained from intensity-based LIF and shadowgraph-based measurements for two initial LiBr mass fractions. Overall, a very good agreement is observed throughout the evaporation process.

A slight delay is systematically observed in the LIF-based measurement, particularly for case a). This discrepancy is likely due to the localized nature of the LIF measurement volume, which probes a thin region near the droplet center, whereas shadowgraph reconstruction integrates the full droplet volume. Since evaporation is typically more intense near the droplet contact line, spatial non-uniformities in concentration may explain this difference.

The theoretical final equilibrium mass fraction $X_{f,theo}$ is also reported. It is obtained by assuming equilibrium between vapor pressures of the ambient humid air $p_{air}$, depending on ambient temperature $T_{amb}$ and relative humidity RH, and the LiBr solution $p_{LiBr}$ at the final droplet temperature, taken equal to the substrate temperature $T_s$ (Eq.9). Both of these pressures are obtained with CoolProp 7.2.0.

$$p_{air}(T_{amb}, RH) = p_{LiBr}(X_{f,theo}, T_S) \qquad \text{(Eq.9)}$$

While good agreement is obtained for case b), a significant deviation is observed for case a). This may be due to imperfect thermal contact between the sapphire substrate and the holder, leading to an underestimation of the wall temperature and so of the theoretical equilibrium. This could also explain why, even if both droplets have similar initial volume, the droplet evaporation duration for case a) is significantly larger than in case b), while having a lower initial salt mass fraction.

Even in this case, a good agreement is still observed between the two independent methods for estimating the LiBr mass fraction, confirming the robustness of the concentration measurement itself.

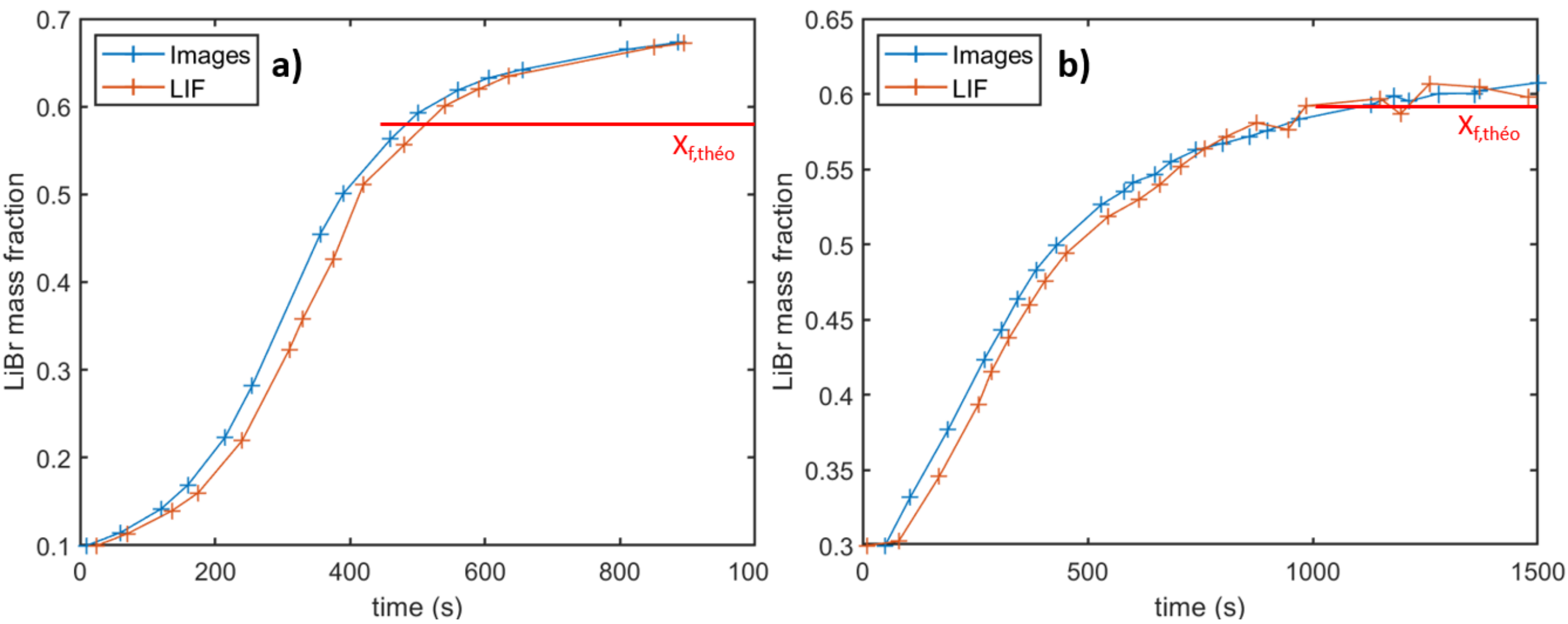

**Figure 14 :** Evolution of the mass fraction of a sessile LiBr droplet evaporating on a heated sapphire at 40 °C

Two general sources of uncertainty were identified from these comparisons:

- First, the progressive increase of dye concentration during evaporation may induce reabsorption effects, particularly due to spectral overlap between absorption and emission bands as represented in annex B. Over the full evaporation process, the dye concentration may increase by up to one order of magnitude. Calibration experiments performed on diluted solutions indicate that this effect leads to an uncertainty below 0.8 wt.% LiBr for an initial dye concentration of $10^{-7}$ M, increasing up to approximately 2 wt.% for higher initial concentrations of $10^{-6}$ M.
- Second, the validation was extended to NaCl droplets (Figure 15). While good agreement is observed during the early and intermediate stages of evaporation, significant deviations appear in the late stage. Unlike LiBr, NaCl droplets reach crystallization conditions, ultimately leading to solid formation. In this regime, shadowgraph-based reconstruction becomes unreliable due to strong morphological changes and optical opacity of the solid phase. In parallel, the LIF-based estimation shows a decrease in the apparent salt mass fraction during the final stage of evaporation. This behavior indicates that the calibration established in the purely liquid phase is no longer strictly applicable once crystallization occurs. The very high local salt concentration, combined with the presence of a solid phase, likely modifies the fluorescence response through enhanced quenching and changes in dye–solute interactions. Experimentally, this regime is characterized by a strong reduction in fluorescence intensity, consistent with a transition beyond the validity domain of the present calibration, which is restricted to liquid-phase saline solutions.

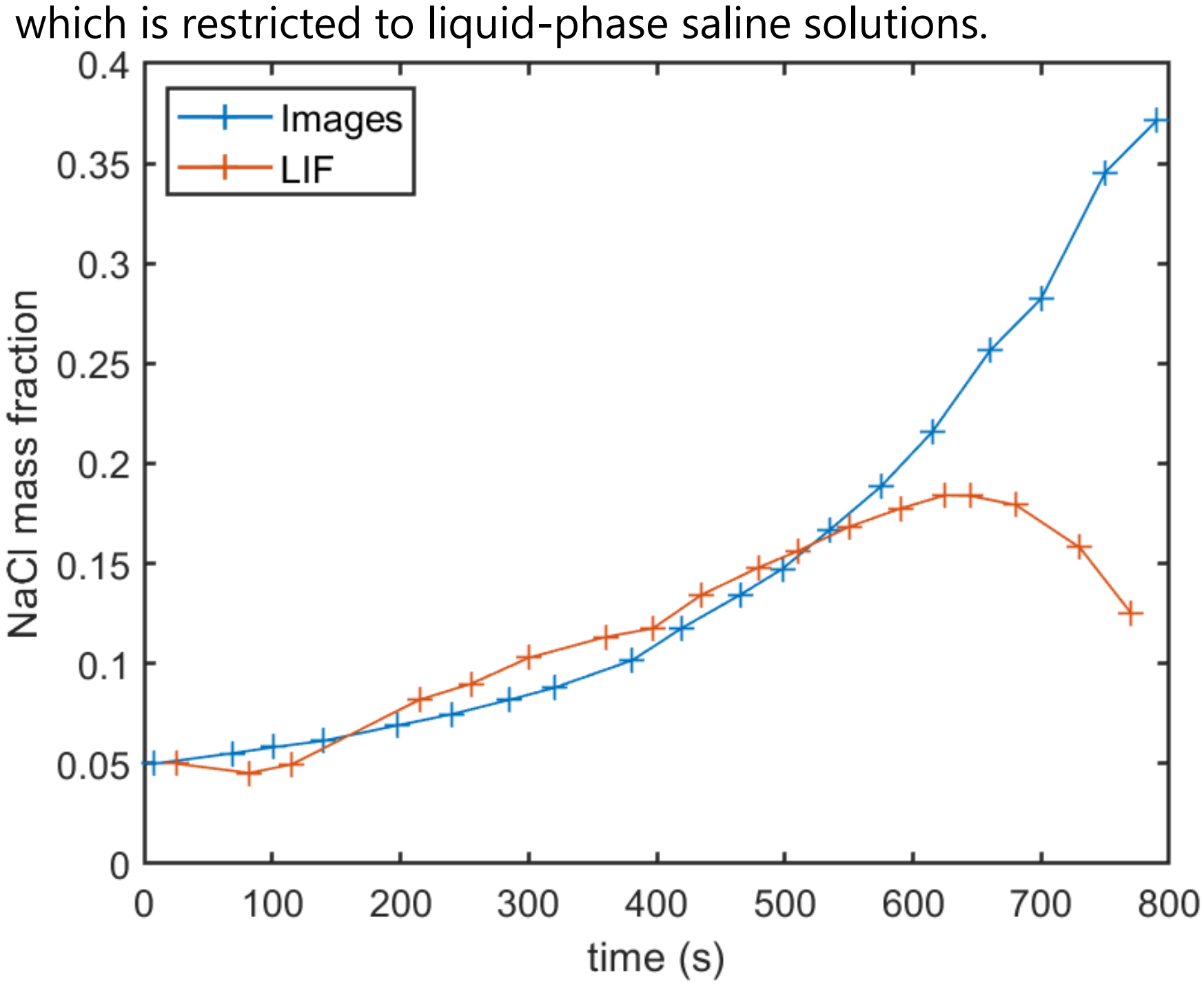

**Figure 15 :** Evolution of the mass fraction of a sessile NaCl droplet evaporating on a heated sapphire at 40 °C

### 4.2 Evaporation of a levitating droplet: lifetime-based LIF and shadowgraph

Finally, validation experiments were performed on acoustically levitated droplets under ambient conditions, using Eosin Y (EY) for fluorescence lifetime measurements, due to its high mass fraction sensitivity and poor-sensitivity to temperature variations. The experimental setup, represented in Figure 16, is similar to the one used for the calibration (Figure 8), the cuvette containing the solution being replaced by the levitator with a droplet of about 20 µL. In this configuration, evaporation occurs in quiescent air (≈ 20 °C and 40 % relative humidity), and proceeds over significantly longer time scales, up to approximately 1.5 hours, due to the absence of external heating.

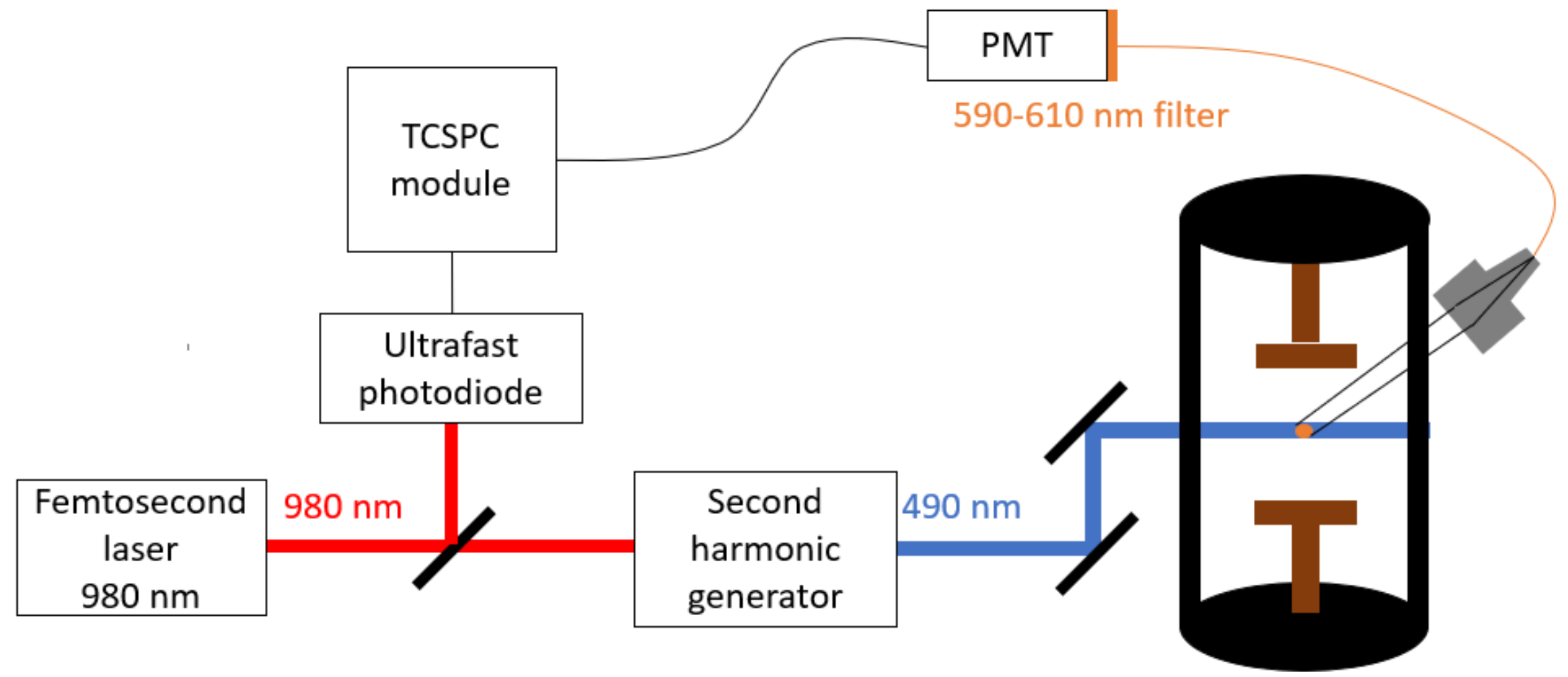

**Figure 16 :** *Experimental TCSPC setup applied on an acoustically levitating droplet*

A typical curve of evaporation of a LiBr levitating droplet is shown in Figure 17, with raw images of the droplet and binarized images after image processing used for measuring the volume variation. It is observed that droplets are slightly compressed, as a result of the acting forces required for maintaining the levitation. The image processing and LiBr mass fraction estimation from volume variation is similar to the procedure described in section 4.1.

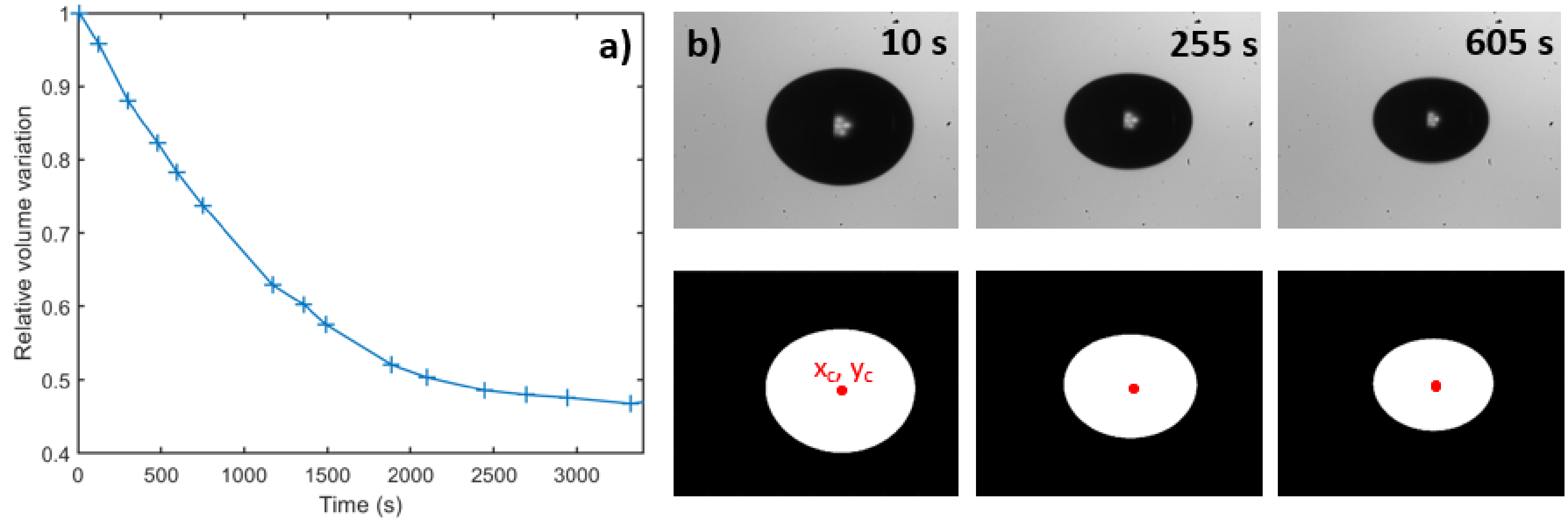

**Figure 17 :** Typical volume reduction of a levitating droplet t a) and images taken (top row), and images after treatment (bottom row). X0=30 %

The comparison between the two techniques is presented in Figure 18, that shows that good agreement is obtained between the two techniques throughout most of the evaporation process, confirming the robustness of the lifetime-based approach under ambient and weekly forced conditions.

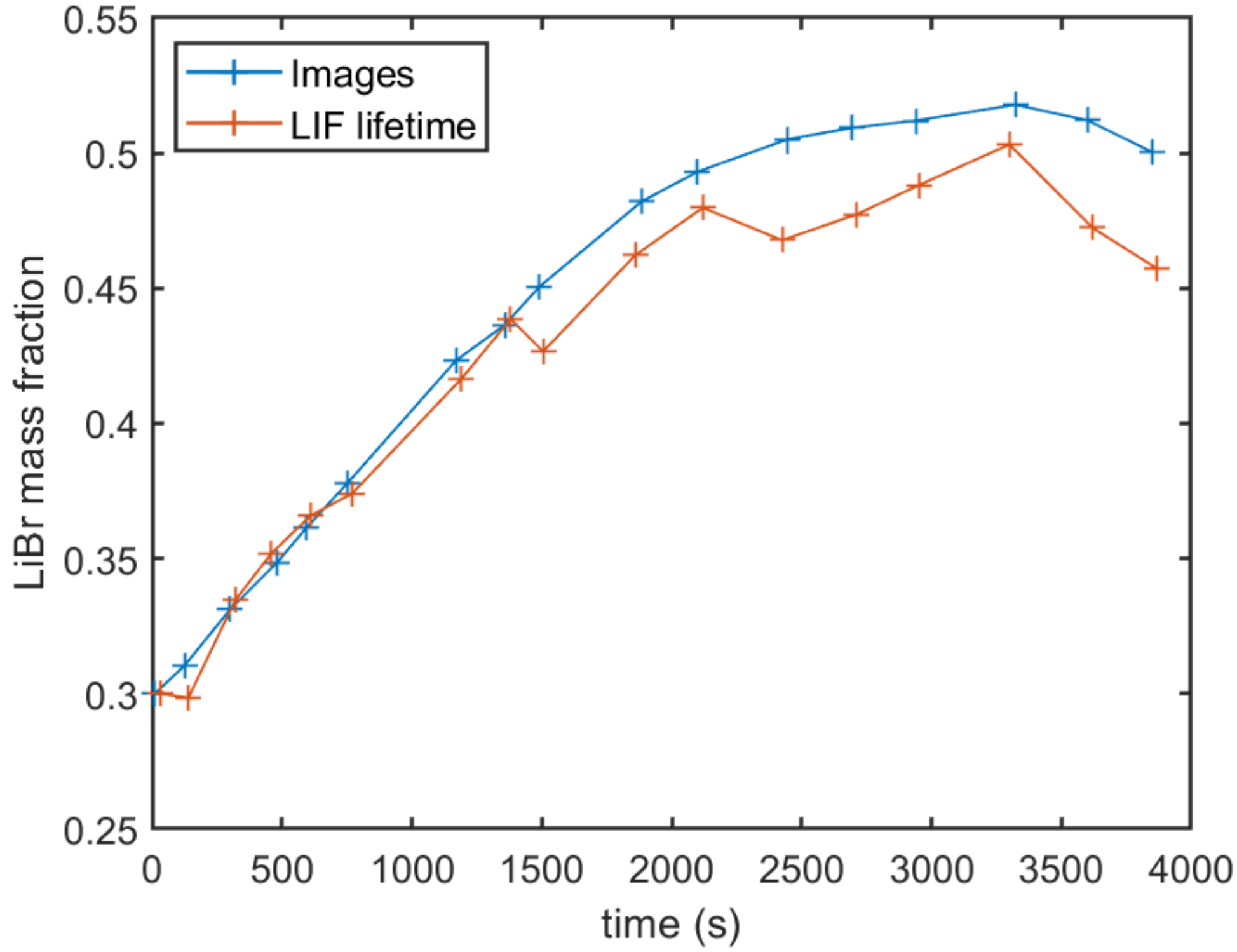

**Figure 18 :** Evolution of the mass fraction of a levitating LiBr droplet

In contrast, the intensity-based ratiometric measurements using KR applied on levitating droplets exhibits a lack of temporal stability. Significant variations in the observed trends are reported between different droplets, even in pure water and sometimes at the early stage of droplet evaporation, preventing a consistent quantitative interpretation. The late stage of evaporation has also demonstrated different trends that suggest that additional slow processes may influence the optical response of the dye. A possible explanation is the development of interfacial and/or photochemically induced effects, such as dye aggregation or modification of dye-solute interactions near the droplet surface. These effects become increasingly relevant over extended time scales and cannot be fully mitigated by adjusting the initial dye concentration within the investigated ranges. These observations highlight that apparently simple configurations, such as acoustically levitated droplets, may in fact involve additional physico-chemical complexity when probed over long time scales, and that the robustness of fluorescence-based diagnostics depend on the chosen observable.

## 5. Concluding remarks and perspectives

A non-intrusive LIF technique has been developed for measuring salt mass fraction in highly concentrated aqueous solutions. The approach is based on the shift of emission spectrum of fluorescent dyes with salt mass fraction towards higher wavelength. A ratiometric formulation using the logarithm of the intensity ratio between two spectral bands enables quantitative concentration measurements with good sensitivity and robustness. The shift has been observed for many dyes and two salt species, namely LiBr and NaCl. The effect of LiBr mass fraction has also been observed on fluorescence lifetime using TCSPC technique.

Both intensity-based ratiometric measurements and lifetime-based measurements have been validated through comparison with independent shadowgraph-based measurements during the evaporation of droplets. The results demonstrated good agreement between the methods. Sources of uncertainty, including temperature effects, reabsorption phenomena, and crystallization-induced phase changes, have been identified and discussed.

Beyond pointwise measurements, the intensity-based LIF approach appears particularly promising for time and spatially resolved diagnostics. Preliminary results on planar imaging of evaporating droplets are currently being investigated and show encouraging potential (Figure 19).

The measurement system will then be extended to study fluctuations in liquid films within desorbers for absorption machines, for which preliminary observations of low-pressure boiling phenomena have already been conducted. For illustrating, typical phenomena observed is shown in Figure 20, in which giant bubbles growth and explosion are projecting droplets and liquid films on a heated surface, leading to intensification of heat and mass transfers. This new LIF technique is expected to be particularly valuable in the context of falling-film exchangers, providing deeper insight into absorption/desorption processes and supporting component optimization.

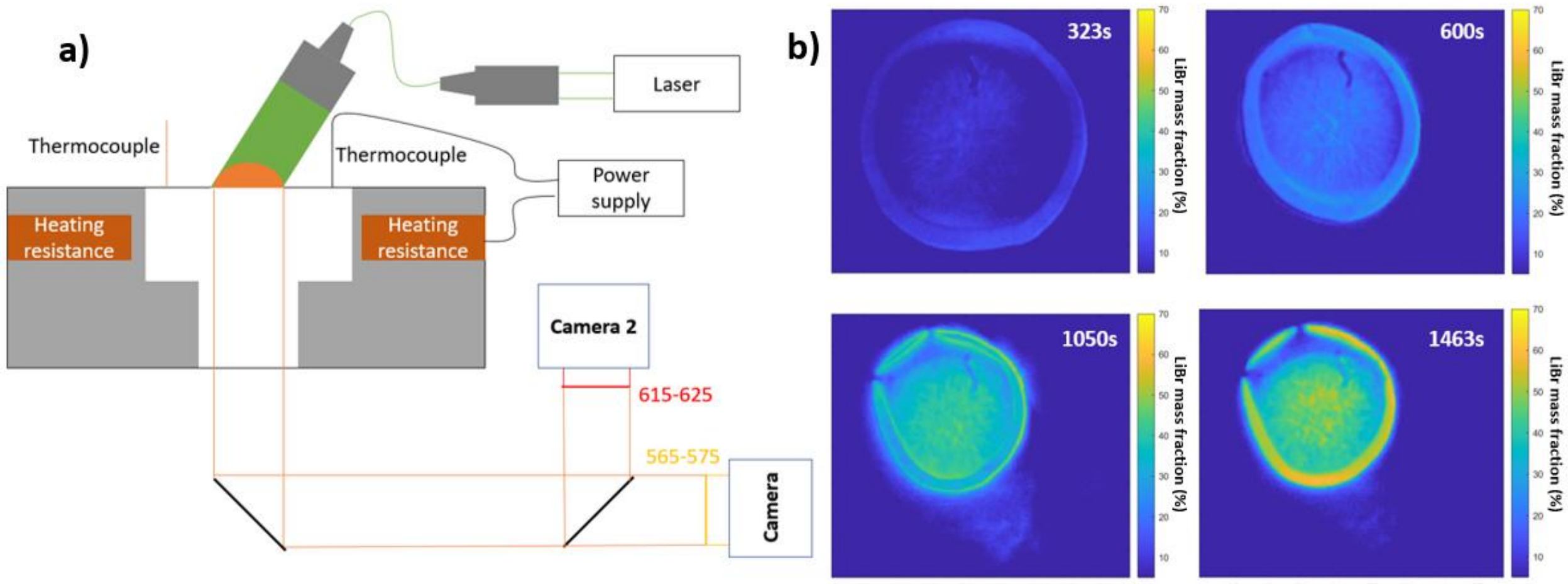

**Figure 19 :** Preliminary tests on imaging applied to the developed LIF technique: a) experimental setup and b) evolution of mass fraction of a heated sessile droplet

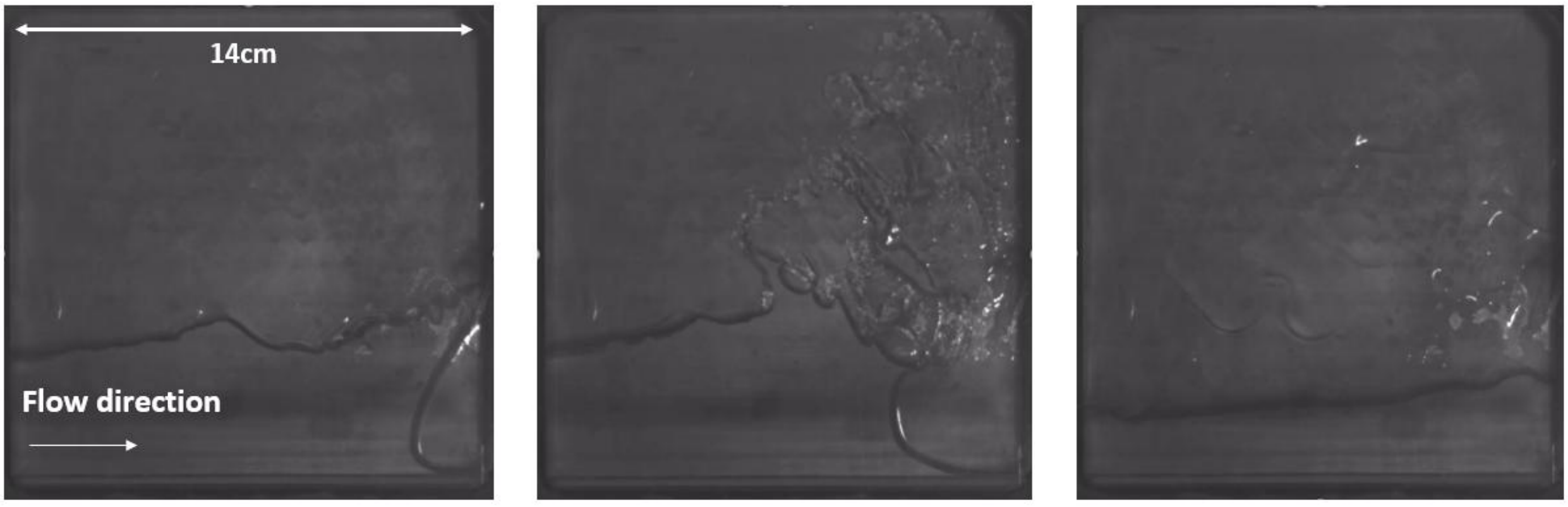

**Figure 20 :** Typical phenomena observed for water boiling at low pressure in a confined channel. Fluid flows from right to left and is heated from the back with a secondary fluid. Chanel thickness is 5.1 mm, absolute pressure is 10 mbar, and secondary fluid temperature is 30 K hotter than saturation temperature ($T_{sat}$ = 7°C at 10 mbar)

## Acknowledgement

Authors would like to thanks Agence Nationale de la Recherche for funding this project (ANR-23-CE05-0001)

## Appendix

### A. Emission spectrum of all tested dyes and salt species

Figures 21-25 represent normalized emission spectrum of dyes from Table 1, and their salt mass fraction sensitivity by wavelength. Most of them present a lower amount of fluorescence light emitted for a given dye concentration or/and spectral conflicts with the laser, leading to a noisy spectrum (especially for RhWt, SRhG and St6). Ox4 exhibits particularly high sensitivity to salt concentration due to the emergence of a secondary fluorescence peak around 660 nm. As the LiBr mass fraction increases, the intensity of this peak grows and becomes predominant at 60 wt.% LiBr. This phenomenon is likely due to the formation of an alternative dye species under high-salinity conditions.

Figure 26 represents the normalized emission spectrum of KR in NaCl solution at 40 °C.

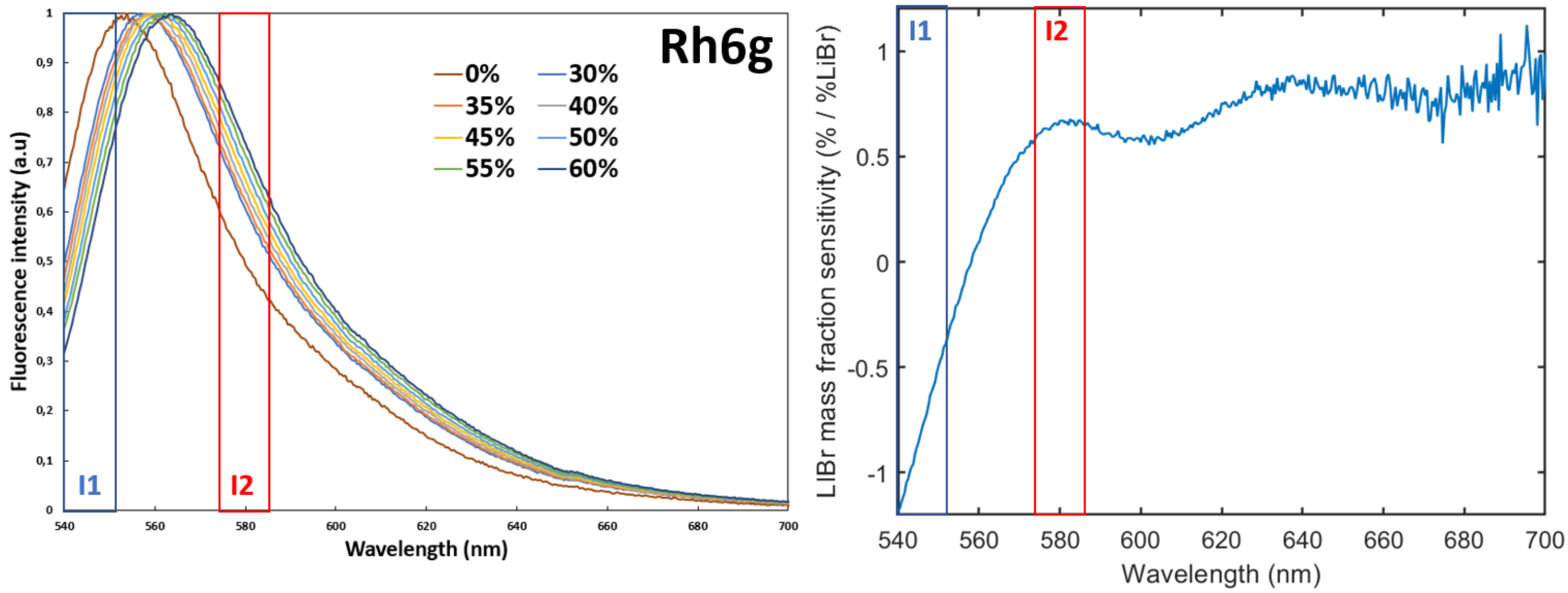

**Figure 21 :** Spectrum of Rh6g in LiBr for various mass fraction, with LiBr sensitivity by wavelength

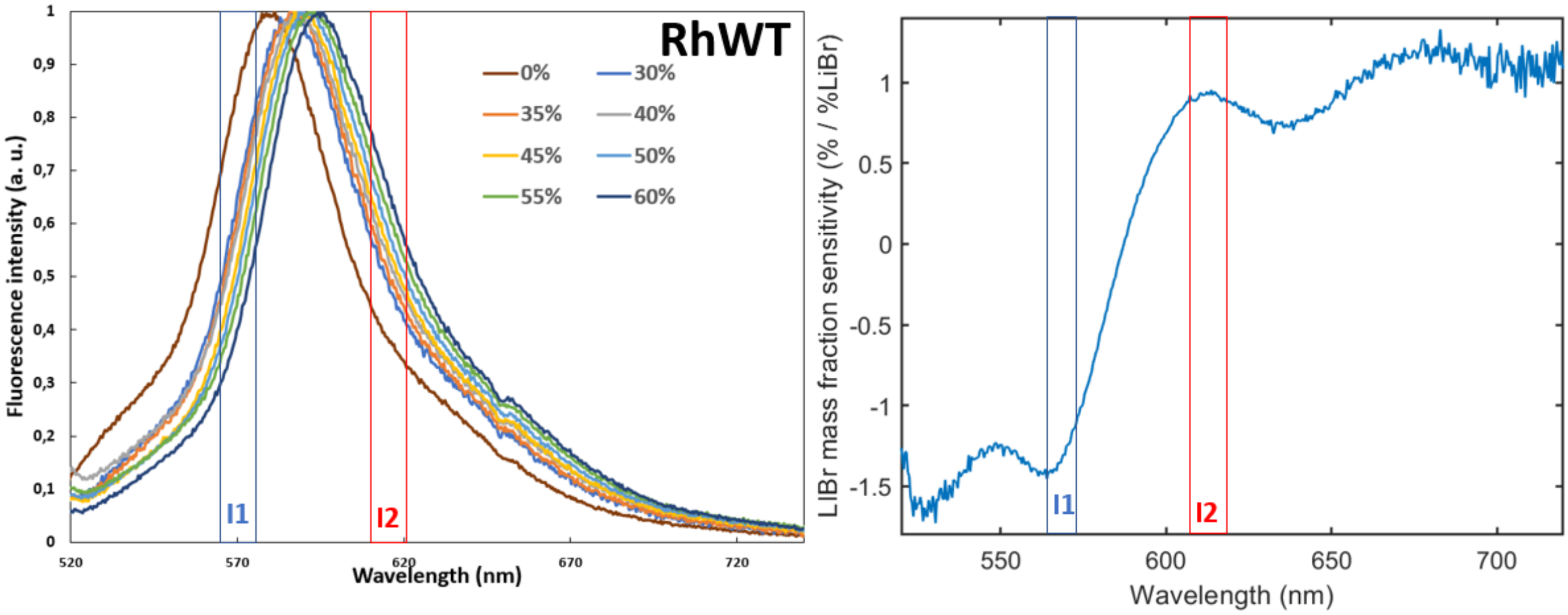

**Figure 22 :** Spectrum of RhWT in LiBr for various mass fraction, with LiBr sensitivity by wavelength

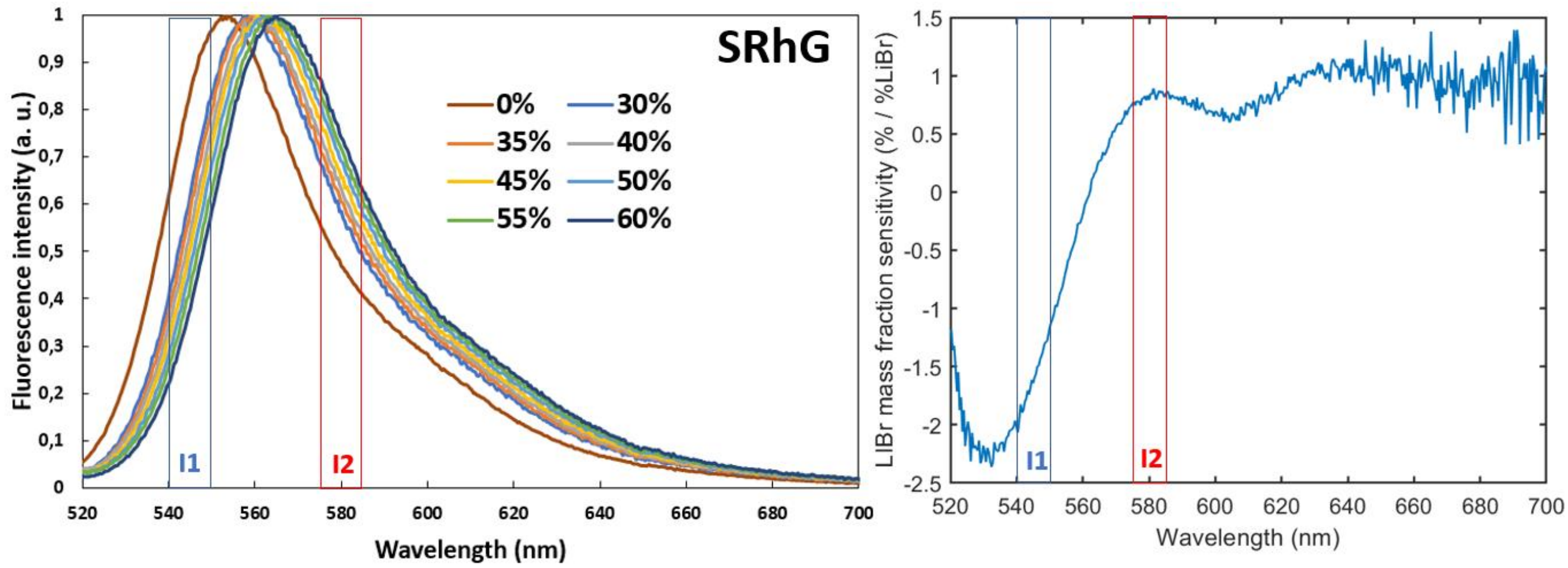

**Figure 23 :** Spectrum of SRhG in LiBr for various mass fraction, with LiBr sensitivity by wavelength

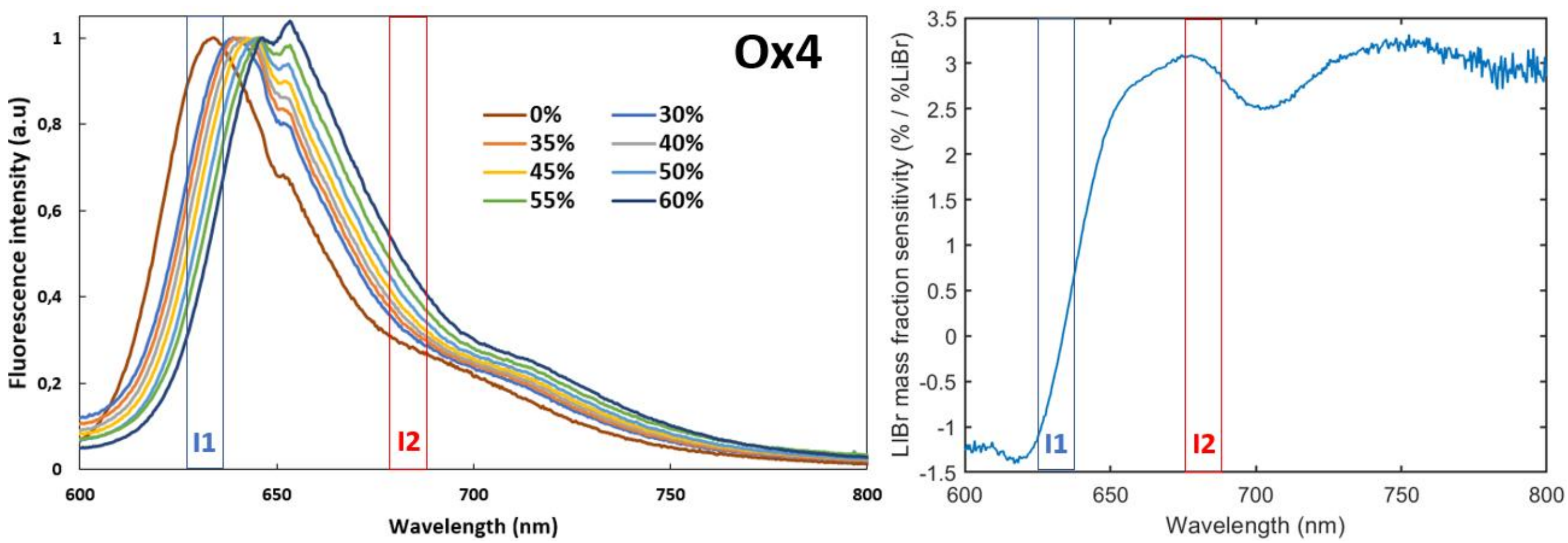

**Figure 24 :** Spectrum of Ox4 in LiBr for various mass fraction, with LiBr sensitivity by wavelength

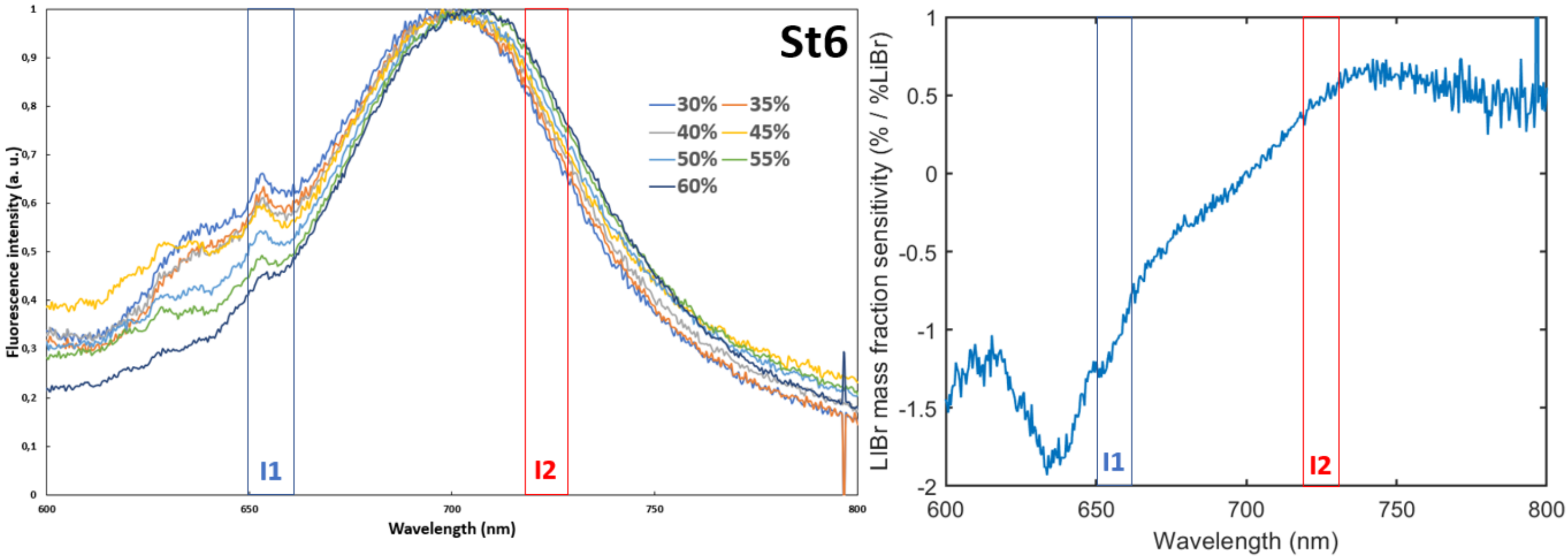

***Figure 25 :*** *Spectrum of St6 in LiBr for various mass fraction, with LiBr sensitivity by wavelength*

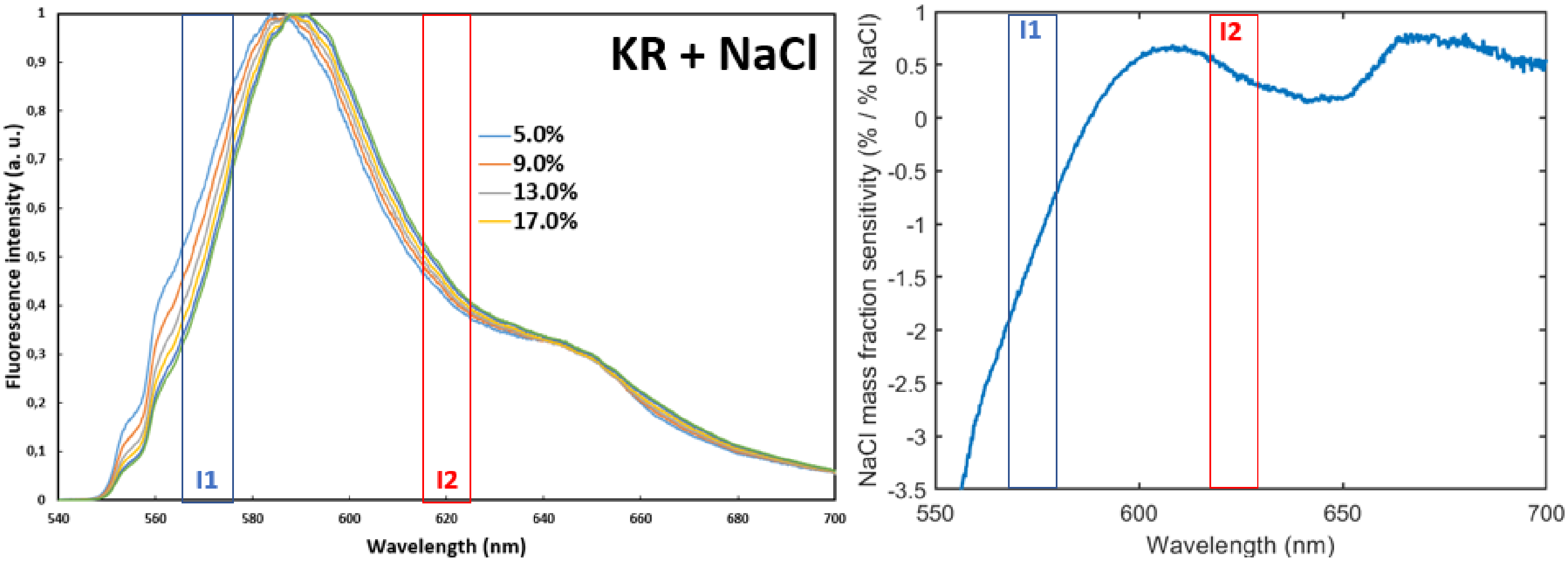

*Figure 26 : Spectrum of KR in NaCl for various mass fraction, with NaCl sensitivity by wavelength taking X1 = 5% and X2 = 25%*

## B. Shift of the absorption spectrum with salinity

Absorption and emission spectra of Kiton Red In LiBr for 0 wt.% and 60 wt.% mass fraction are represented In Figure 27. It is observed that both spectra are shifting in a similar way with LiBr mass fraction. It is also noted that emission and absorption spectra overlap each other, potentially causing reabsorption, as discussed in section 4.1.

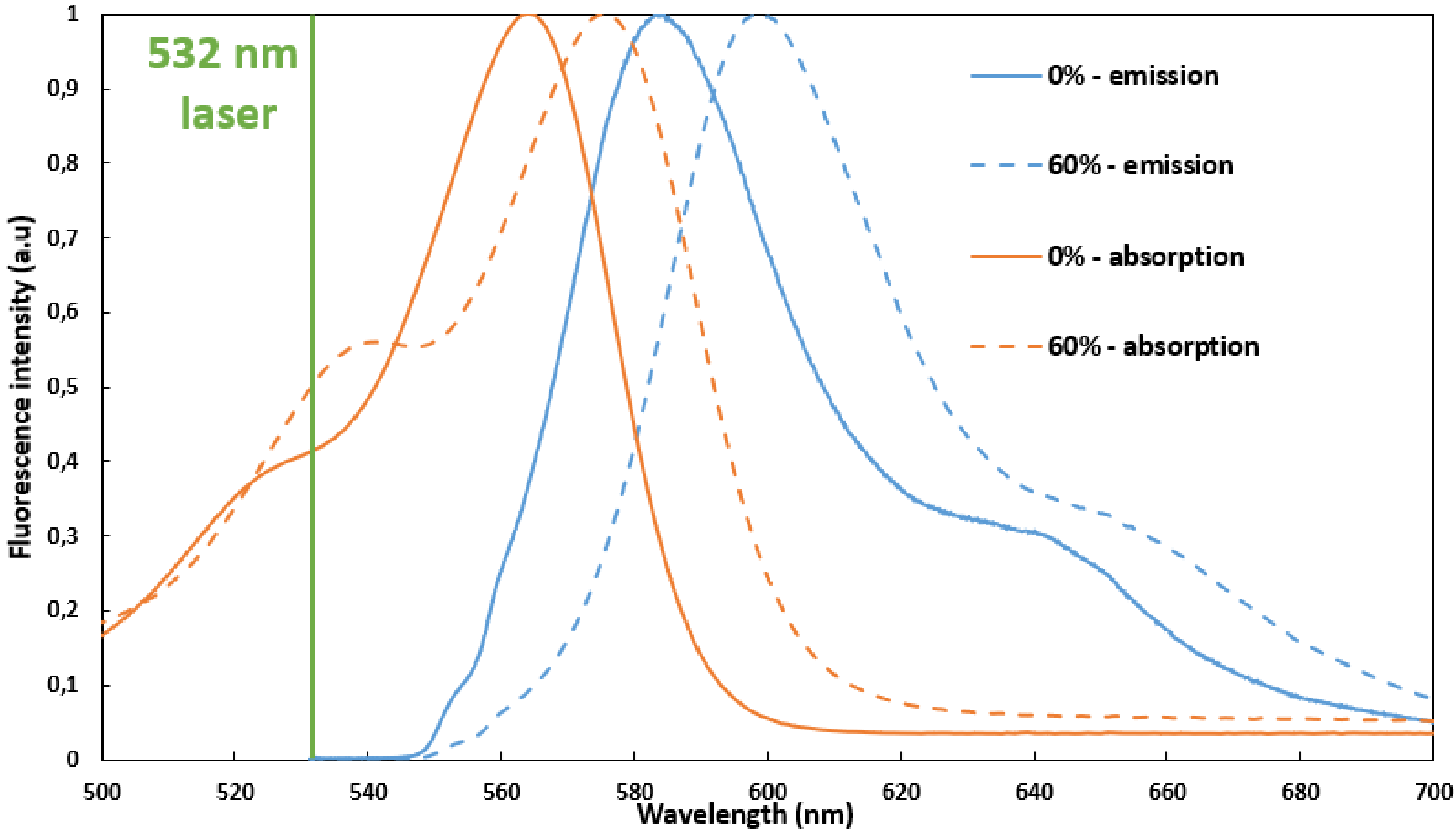

**Figure 27 :** Comparison of absorption and emission spectra of Kiton Red for 0 wt.% and 60 wt.% LiBr mass fraction